# The N$_2$V color center: a ubiquitous visible and near-infrared-II quantum emitter in nitrogen-doped diamond

Brett C. Johnson[1], Mitchell O. de Vries[1], Alexander J. Healey[1], Marco Capelli[1], Anjay Manian[1], Giannis Thalassinos[1], Amanda N. Abraham[1], Harini Hapuarachchi[1], Tingpeng Luo[2] Vadym Mochalin[3], Jan Jeske[2], Jared H. Cole[1], Salvy Russo[1], Brant C. Gibson[1], Alastair Stacey[1], Philipp Reineck[1]

[1] School of Science, RMIT University, Melbourne, VIC 3001, Australia
[2] Fraunhofer Institute for Applied Solid State Physics IAF, Tullastraße 72, 79108 Freiburg im Breisgau, Germany
[3] Department of Chemistry & Department of Materials Science & Engineering, Missouri University of Science & Technology, Rolla, Missouri 65409, USA

**Abstract**
Photoluminescent defects in diamond, like the nitrogen-vacancy (NV) color center, are at the forefront of emerging optical quantum technologies. Most emit in the visible and near-infrared spectral region below 1000 nm (NIR-I), limiting their applications in photonics, fiber communications, and biology. Here, we show that the nitrogen-vacancy-nitrogen (N$_2$V) center, which emits in the visible and near-infrared-II (NIR-II, 1000-1700 nm), is ubiquitous in as-synthesized and processed nitrogen-doped diamond from bulk samples to nanoparticles. We demonstrate that N$_2$V is also present in commercially available state-of-the-art NV diamond sensing chips made via chemical vapor deposition (CVD). In high-pressure high-temperature (HPHT) diamonds, the photoluminescence (PL) intensity of both N$_2$V charge states, N$_2$V$^0$ in the visible and N$_2$V$^-$ in the NIR-II, increases with increasing substitutional nitrogen concentration. We determine the PL lifetime of N$_2$V$^-$ to be 0.3 ns and compare a quantum optical and density functional theory model of the N$_2$V$^-$ with experimental PL spectra. Finally, we show that detonation nanodiamonds (DND) show stable PL in the NIR-II, which we attribute to the N$_2$V color center, and use this NIR-II PL to image DNDs inside skin cells. Our results will contribute to the scientific and technological exploration and development of the N$_2$V color center and inspire more research into its effect on other color centers in diamond.

**Introduction.**
Photoluminescent color centers in diamond have attracted considerable attention in the scientific community over the past two decades for applications from solid-state single photon emission for quantum communication[1] and computing[2] to quantum sensing in biology[3,4]. The most known and technologically developed color center, the nitrogen-vacancy (NV) center, is extensively used as a powerful quantum sensor in fundamental and applied science and is at the heart of many emerging technologies.

The nitrogen-vacancy-nitrogen (N$_2$V) is a closely related color center first observed in absorption measurements in 1956[5]. Its atomic structure was identified in 1976[6] and consists of a pair of nitrogen atoms separated by a vacancy. Like the NV center, it typically exists in a neutral (N$_2$V$^0$) or negative (N$_2$V$^-$) charge state, and the latter was only unambiguously identified as the negative charge state of the N$_2$V in 2015.[7] Unlike the NV, which exhibits overlapping photoemission for both charge states, N$_2$V$^0$ and N$_2$V$^-$ photoluminescence (PL) are spectrally separated by more than 300 nm; at room temperature, the N$_2$V$^0$ emission peak is centered in the green part of the visible spectrum (~540 nm), while the N$_2$V$^-$ emits from 950 nm to 1400 nm in the second near-infrared (NIR-II) spectral region. Very few color centers with PL in the NIR-II have been reported.[8] N$_2$V$^0$ and N$_2$V$^-$ are also commonly referred to as H3 and H2 in the literature, respectively.[9]

Single photon emission has been observed for N$_2$V$^0$ [10] but not for N$_2$V$^-$. A radiative PL lifetime (i.e. in the absence of any non-radiative decay) of 17.5 ns and a PL quantum yield of 0.95 have been reported for N$_2$V$^0$.[11] Neither the PL lifetime nor quantum yield have been reported for N$_2$V$^-$. N$_2$V$^0$ is diamagnetic and has been proposed for quantum memory applications[12] and as a qubit.[13] $N_2V^-$ is paramagnetic (with spin S=½), and its electron paramagnetic resonance (EPR) spectrum has been identified experimentally.[7] $N_2V$ is also known to be photochromic, i.e., light excitation leads to a redistribution of charges and modulates the $N_2V$ charge state.[14] Generally, the charge state of color centers in diamond is determined by the concentrations of electron donors and acceptors and their relative charge state transition levels.[15] Nitrogen is one of the most important electron donors in diamond. Due to its high charge state transition ($N_s^0/N_s^+$) level of ~3.6 eV above the valance band minimum (VBM) and abundance in many diamond samples, it can donate electrons to color centers like the NV to create $NV^-$.[16] However, the spatial distribution of defects in the diamond lattice is also known to play an important role in transitions between different charge states of NV centers.[17] Despite its interesting photoluminescence properties and potential for emerging quantum applications, the $N_2V$ has been studied far less than the NV center. One reason is that the controlled and selective creation of $N_2V$ centers is experimentally challenging and competes with NV center formation. It generally requires extreme processing of HPHT diamonds that usually incorporate substitutional nitrogen at concentrations on the order of 100 ppm from the atmosphere. $N_2V$ creation typically involves irradiation with high energy electrons (>2 MeV) to create vacancies and annealing at high temperatures (above 1400 °C) and pressures (above 4 MPa) to enable vacancy, carbon interstitial, and nitrogen migration to form $N_2V$ (see Ashfold et al.[9] for more details). Recently, Cas et al. showed that high-temperature rapid annealing may be a route to control the relative concentration of several nitrogen-related color centers, including $N_2V$.[18]

Amongst the studies that have investigated the optical properties of $N_2V$, the vast majority focus on its neutral charge state ($N_2V^0$). Few studies have experimentally investigated the photoluminescence properties of $N_2V^-$ [7,14,19,20] and $N_2V^-$ PL from micro- and nanoparticles has not been reported. At the same time, the $N_2V^-$ emits in a spectral region that makes it relevant to both biological and telecommunications applications: in the so-called near-infrared II biological transparency window (1000-1400 nm) and the telecom O-band (1260- 1360 nm), where only one other color center is known to emit.[8] $N_2V^-$ PL can penetrate deep into biological tissue, is compatible with established silicon photonics and optical fiber communications technologies, and thus has technological innovation potential.

The photostability of color centers can be severely compromised by the proximity of the surface. This is most clearly demonstrated in diamonds with nano-scale dimensions.[21] Detonation nanodiamonds (DNDs) are among the smallest nanodiamonds that can be reliably synthesized at scale[22] and have a typical size of 5 nm. Due to their unique physico-chemical properties, they have been extensively studied for their potential in biomedical applications, are readily taken up by many cell types, and are not cytotoxic to many cells.[23] Yet, the efficient creation of color centers with stable PL has proven difficult in DNDs thus far. As a result, few studies have explored the potential of DNDs for biomedical imaging and theranostics. The studies that do exist focus on the visible and NIR-I spectral regions.[23,24] However, particularly for theranostic applications, NIR-II imaging would enable transdermal imaging from greater depths, which is very attractive, especially for applications in medicine, as well as for some animal models.

Here, we demonstrate that the $N_2V$ color center is present in nitrogen-doped diamonds as-synthesized and in standard processed samples containing high densities of NV centers. This includes diamonds made via HPHT synthesis, chemical vapor deposition (CVD), and detonation synthesis (i.e. DNDs). PL from $N_2V^0$ and $N_2V^-$ is found in commercial HPHT diamond from bulk samples to sub-micron particles. We show that $N_2V^0$ and $N_2V^-$ PL increase with increasing substitutional nitrogen concentration and that both $N_2V$ charge states can locally coexist with NV color centers. $N_2V^0$ PL is found in bulk diamonds made via CVD with substitutional nitrogen above 10 ppm. We investigate the $N_2V^-$ PL at cryogenic temperatures (5 K) and determine the $N_2V^-$ PL lifetime as 0.3 ns. The $N_2V^-$ emission properties are modelled using a phenomenological quantum optical approach, calculated using density functional theory (DFT) and compared to experimental results. Finally, we show that

DNDs show strong and stable PL above 1000 nm, which we attribute to the N$_2$V color center. We demonstrate PL imaging in the NIR-II biological transparency window with DNDs taken up by skin cells. Our results pave the way for the scientific and technological exploration and development of the N$_2$V color center in diamond using existing commercially available and newly engineered diamond materials. The ubiquity of N$_2$V in nitrogen-doped diamond, including high-performance NV sensing chips, makes further research into its photophysical properties important for optimizing all optical diamond technologies.

**Results**

We first investigated the photoluminescence (PL) of commercial HPHT diamonds of different form factors from bulk to 1 µm particles using custom-built confocal PL microscopes. All samples were studied as received. Particles were suspended in deionized water and drop-cast onto quartz substrates for PL imaging and spectroscopy. See Methods and SI for more details on microscopes and samples. Figures 1 a) to d) show NIR-II PL images of a bulk diamond sample and 10, 5, and 1 µm diamond particles acquired using 785 nm excitation (1 mW) and detecting fluorescence above 900 nm and mainly in the NIR-II. The bulk sample ((100) HPHT SC plate, Element Six, Figure 1a) shows strong and homogenous PL across the field of view. The PL intensity of the brighter HPHT particles (MSY series, Pureon, Figures 1b)-1d) is of the same order of magnitude as the bulk sample but the PL intensity varies significantly between individual particles. For example, the PL intensity of the brighter 5 µm particles is about 10 times brighter than that of the dimmer particles that are barely visible.

We also investigated air-oxidized detonation nanodiamonds (DND, UD98, NanoBlox) with an average aggregate size of ~ 70 nm using the abovementioned imaging conditions. Figure 1e) shows a typical NIR-II PL image of the DNDs. While the PL intensity of the brighter particles in the image is about 3 times lower than that of the 1 µm HPHT particles, individual particles or small particle aggregates can be identified. PL from all samples, including DNDs, was extremely photostable over minutes of continuous photoexcitation, an important and characteristic feature of diamond color centers.

Figure 1 f) shows typical room temperature PL spectra for the samples shown in Figure 1 images a)-e) acquired using 470 nm and 785 nm laser excitation for the visible and NIR-II spectral regions, respectively. Upon 470 nm excitation, all samples except DNDs show broad emission from 500 to 700 nm, peaking at ~ 530 nm, and a small narrow peak at 503 nm, characteristic of the neutral charge state of the N2V color center (Figure 1 f, left).[25] The inset on the left of Figure 1 f) shows a zoom into the so-called zero-phonon line (ZPL) peak at 503 nm. The ZPL is most pronounced for the bulk sample and broadens with decreasing particle size. DNDs show weak PL in the visible that is not photostable and likely originates from residual non-diamond carbon, as discussed in detail elsewhere.[26]

When excited with 785 nm light, all HPHT diamond samples show PL from 900 to 1400 nm with a broad peak centered at ~1120 nm and a spectrally narrower peak at 986 nm, characteristic of the ZPL of N$_2$V$^-$ (Figure 1 f, right).[5] The ZPL is most pronounced for the bulk samples and becomes less pronounced and broader (6.5 ± 0.2 nm FWHM for bulk to 10.5 ± 1.8 nm for 1 µm particles). This will be investigated in more detail in Figure 4. The DND PL spectrum has a similar spectral shape as the spectra of the HPHT particles but does not exhibit a ZPL peak at 986 nm. Our data strongly suggests that the PL originates from the N$_2$V$^-$ color center, which will be discussed in more detail below.

Figure 1 shows that both charge states of the N$_2$V color center are present in unprocessed HPHT diamond bulk and particle samples down to 1 µm particle size. To demonstrate the ubiquity of N$_2$V in HPHT diamonds, we also investigated bulk and particle samples from other suppliers as shown in SI Figures S1-S4. We find that HPHT bulk samples from three different suppliers and particle samples (at least three different particle sizes from 0.5 µm to 10 µm) from two suppliers all exhibit the spectral PL features characteristic of N$_2$V$^0$ and N$_2$V$^-$ shown in Figure 1 f).

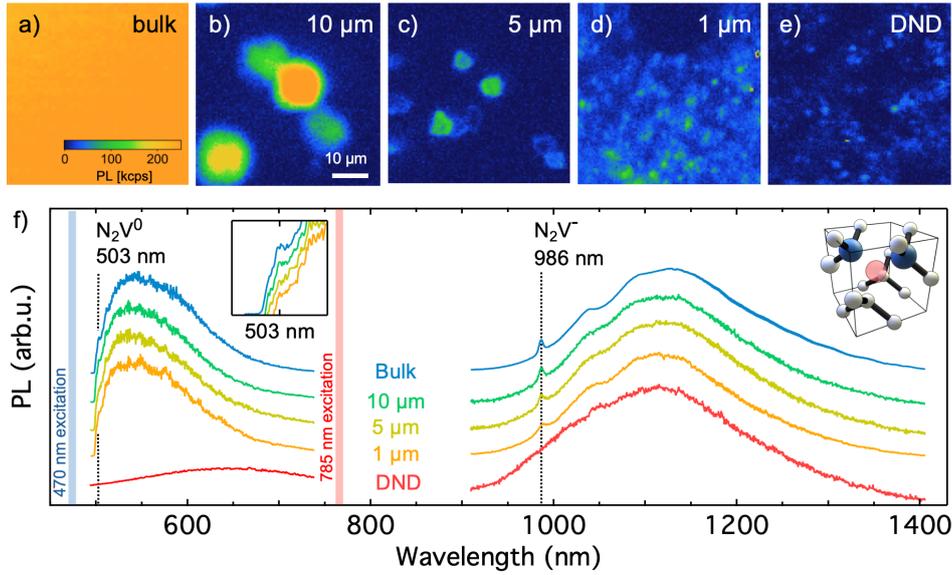

*Figure 1. Visible and NIR-II PL from $N_2V^0$ and $N_2V^-$, respectively. a)-e): NIR-II PL images of bulk HPHT diamond (a), 10 µm (b), 5 µm (c), 1 µm (d) HPHT particles and detonation nanodiamonds (DND) (e). Images were acquired in the 900-1500 nm spectral region using 785 nm excitation. The false color PL intensity scale in a) and the scale bar in b) apply to all images. PL intensity is in thousand photocounts per second (kcps). f) Typical $N_2V^0$ and $N_2V^-$ spectra of samples shown in panels a) to e) using 470 nm ($N_2V^0$) and 785 nm ($N_2V^-$) excitation. The inset on the left side of f) shows a zoom into the ZPL region of the $N_2V^0$ spectra. The inset in the top right corner shows the atomic structure of $N_2V$ where grey and blue spheres are carbon and nitrogen atoms, respectively, and the pale red sphere represents a vacancy.*

The presence of nitrogen in the diamond lattice is essential for the formation of $N_2V$. In HPHT diamonds, nitrogen is present primarily either in the form of single nitrogen atoms replacing a carbon atom ($N_s^0$, also called P1-center) or in the form of two nitrogen atoms substituting two neighboring carbon atoms ($N_2$ or A-center).[27] $N_s^0$ is the most common impurity and is known to be present at concentrations of ~100 ppm. Therefore, we investigated the relationship between the $N_s^0$ concentration and the PL intensity of $N_2V$. The $N_s^0$ concentration strongly depends on the exact synthesis conditions, precursors, and environment. We studied several bulk HPHT samples from three suppliers to cover a significant $N_s^0$ concentration range between 100-400 ppm.

Figure 2 a) and b) show the integrated PL intensity of $N_2V^0$ and $N_2V^-$, respectively, as a function of $N_s^0$ concentration in (100) diamond plates. For each sample, the $N_s^0$ concentration was determined based on the known infrared absorption of $N_s^0$ [28] in six locations on the sample using Fourier transform infrared (FTIR) absorption spectroscopy (see Methods and SI Figure S5 for details) and averaged. The average values are plotted in Figure 2 a) and b), and the horizontal error bar represents the standard deviation between individual measurements. A 200 × 200 µm PL image was acquired for each sample, and the total PL intensity in the brightest and dimmest region of the image was determined. The average total PL intensity from these two regions is plotted in Figures 2 a) and b), and the vertical error bars represent the minimum and maximum values.

The $N_2V^0$ and $N_2V^-$ PL increase monotonously with increasing $N_s^0$ concentration in the range between 130 and 350 ppm. The solid lines in Figures a) and b) are linear fits to the experimental data, and a are to the eye only. While the number of data points and range of $N_s^0$ concentration investigated here are insufficient to establish a functional relationship, Figures 2 a) and b) show that the PL of both $N_2V$ charge states increases with increasing $N_s^0$ concentration in HPHT diamond samples. The PL quantum yield (0.95[11]) and absorption cross-section ($2.1 \times 10^{-17}$ cm$^2$ [29]) are known for $N_2V^0$ but not for $N_2V^-$. By comparing the PL intensity of $N_2V^0$ to that of an NV reference sample with a known NV concentration, we estimate the absolute concentration of $N_2V^0$ centers (Figure 2 a, right y-axis) to increase from 1 to 11 ppb as the $N_s^0$ concentration increases from 130 to 350 ppm. This would suggest that only ~1 $N_2V^0$ is present for every 130,000 $N_s^0$. However, a much higher relative concentration of $N_2V^-$ might be present, as will be discussed in more detail below. Interestingly, we find that only the unprocessed bulk sample with the highest $N_s^0$ concentration of 350 ppm shows NV PL upon 520 nm light excitation, suggesting that the formation of $N_2V$ is not correlated with NV formation (see SI Figure S6). This will be investigated further in the next section.

We now consider growth sectors neighboring the main (100) sector. Figure 2 c) shows a schematic illustration of a typical (100) diamond plate and a zoom-in to one of the edge regions containing other growth sectors. A confocal PL image (470 nm excitation, > 490 nm PL collection) of the edge region, as indicated in Figure 2 c), is shown in Figure 2 d). While the (100) sector shows bright $N_2V^0$ PL, the neighboring sectors show no PL at all (middle, (113) sector) or weak $NV^-$ PL (bottom left, (111) sector). Similarly, upon 785 nm excitation, the (100) sector shows strong $N_2V^-$ PL while the two other regions show no PL in the NIR-II (see SI Figure S7). In HPHT diamond, the $N_s$ incorporation probabilities in different growth sectors typically follow the trend: (111) > (100) > (113).[30] For the HPHT sample investigated here, our results suggest that the $N_s^0$ concentration is highest in the (100) sector (see SI Figure S5). $N_2^0$ (A-center) could not be identified based on its characteristic IR absorption at 1282 cm$^{-1}$ in any of our FTIR spectra, suggesting that nitrogen is mainly present in the form of $N_s$.

Nitrogen plays a double role: as a building block for $N_2V$ and as an electron donor that governs the charge state of color centers in its vicinity. Figure 2 e) shows the charge transition levels of $N_s$, $N_2V$ and NV calculated using DFT.[15] The relatively high transition level of $N_s^{0/+}$ (~3.6 eV) compared to that of $N_2V^{-/0}$ (~3.2 eV) means that $N_s^0$ can donate an electron to nearby $N_2V^0$ defects to create $N_2V^-$ and $N_s^+$. Our relative concentration estimate for $N_2V^0$ to $N_s^0$ of 1 to 130,000 suggests a large excess of $N_s^0$ electron donors that are available for every $N_2V$ defect. Considering only the transition levels shown in Figure 2 e), this would mean that all $N_2V$ emitters are in their negative charge state. The fact that $N_2V^0$ PL is observed in all samples and that this PL intensity follows the same trend as $N_2V^-$ suggests that other factors, such as the spatial distribution of defects in the crystal,[17] limit the complete conversion of $N_2V^0$ to $N_2V^-$. It also indicates that, in general, the concentration of $N_2V^-$ might be significantly higher than that of $N_2V^0$.

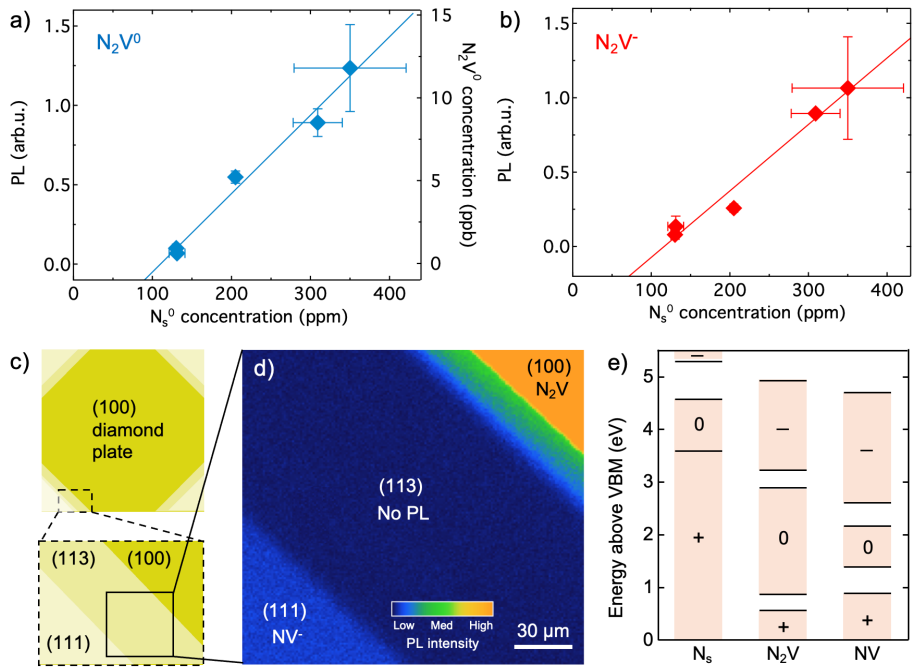

*Figure 2. a) & b) $N_2V^0$ (a) and $N_2V^-$ (b) PL intensity as a function of substitutional nitrogen ($N_s^0$) concentration in the (100) growth sector of bulk HPHT samples. In both graphs, the solid line is a linear fit to the data points. The $N_s^0$ concentration was determined as averages from FTIR spectra recorded in six locations. Horizontal error bars represent the standard deviation of these measurements. Vertical error bars represent the range of PL intensities observed across a 200 × 200*

µm area in the center of the sample. c) Schematic illustration of a (100) HPHT diamond plate as seen from the top. Lighter-shaded areas in the corners indicate different growth sectors, one of which is enlarged in the bottom inset d) Confocal PL image (470 nm excitation, PL 490-900 nm detected) of the part of a (100) diamond sample illustrated in panel c) that shows three growth sectors: (100) (top right), (113) (middle) and (111) (bottom left). e) DFT charge state transition levels of $N_s$, $N_2V$ and $NV$[15].

This donor ($N_s^+$)/acceptor ($N_2V^-$) relationship is analogous to $N_s^+$ and $NV^-$ and is known to be the main mechanism for creating $NV^-$ color centers in highly nitrogen-doped diamond samples.[16] The even lower transition level of $NV^{-/0}$ relative to $N_2V^{-/0}$ suggests that $N_2V^-$ can also donate an electron to nearby NV centers and raises the question of whether $NV^-$ and $N_2V^-$ can locally co-exist and produce PL in the diamond crystal lattice. To address this question and identify if NV and $N_2V$ PL are correlated more generally, we investigated 5 µm HPHT particles by selectively exciting these different color centers. Figures 3 a) to c) show PL images of the same particles acquired using 470, 520, and 785 nm lasers to excite $N_2V^0$, NV, and $N_2V^-$, respectively, and collecting PL in the spectral regions indicated in Figure 3 g). Most particles are visible in all three spectral channels, demonstrating that $N_2V^0$, $NV^-$, and $N_2V^-$ are all present in these particles. However, the relative brightness of each defect and charge state varies greatly, and many particles show only NV PL and no $N_2V^{0/-}$ PL and vice versa (see SI Figure S8).

The fact that NV and $N_2V$ centers can locally coexist suggests that the two defects can exchange electrons via photoionization processes analogous to those known to occur between NV and $N_s$. For NV-based technologies, this is relevant for two reasons: 1. The presence of $N_2V$ may create additional non-radiative decay pathways for $NV^-$ centers via electron transfer from excited state $NV^-$ to nearby $N_2V$ defects and thus reduce the PL quantum yield of $NV^-$. 2. $N_2V^-$ shows strong absorption in the $NV^-$ spectral emission band around 700 nm and may re-absorb photons emitted by NV centers. Induced absorption in the $NV^-$ emission region caused by green laser pumping has been observed in optical cavity experiments[31] and has high relevance for laser threshold magnetometry.[32,33] The creation of $N_2V^-$ via photoionization is a possible explanation for this induced absorption.

Smaller HPHT diamond crystals like those in Figures 3 a) to c) are usually milled from larger crystals. Therefore, the 5 µm particles investigated here can originate from different growth sectors and are expected to display broad variations in $N_s^0$ concentrations. To investigate if the $N_2V$ PL brightness of the particles also increases with increasing $N_s^0$ concentration, as seen for bulk samples in Figure 2, we performed optically detected magnetic resonance (ODMR) measurements. Using wide-field microscopy,[34] ODMR spectra (Figure 3e) were recorded for each pixel in the region shown in Figures 3 a) to c), and the experimental data (circles) were fitted with a double Lorentzian (solid line). Figure 3 d) shows a map of the full-width half maximum (FWHM) of one of the Lorentzians. The white circles indicate two particles with very high (bottom circle) and low (top circle) $N_2V$ PL from both charge states. The ODMR spectra of these two particles are plotted in Figure 3 e) and show an FWHM (one of the two Lorentzians) of 7 MHz for the dimmer particle (top) and 17 MHz for the bright particle (bottom). The ODMR linewidth is known to increase with increasing $N_s^0$ concentration by 10 MHz per 100 ppm of $N_s^0$,[35] and hence we expect nitrogen to be the dominant source of line broadening in these samples. This suggests that the brighter particle has ~100 ppm more $N_s^0$ than the dimmer particle, which agrees with the trend observed in Figure 2 b). We determined the ODMR width for 16 particles in total (see SI Figure S9 for the entire ODMR map analyzed). Figure 3 f) shows the $N_2V^-$ PL intensity of those particles as a function of their ODMR width and a linear fit to the experimental data. A statistical rank analysis reveals a strong positive correlation between $N_2V^-$ PL and ODMR width (Spearman coefficient $r_s$=0.72), suggesting that $N_2V^-$ PL intensity scales with $N_s^0$ concentration in 5 µm HPHT particles.

We then further investigated if the $N_2V$ PL intensity correlates with the NV PL intensity. Figure 3 h) shows normalized $N_2V^-$ PL as a function of $NV^-$ PL determined for 53 individual particles (see SI Figure S10 for the investigated PL images). The scatter plot shows that the $N_2V^-$ and $NV^-$ intensities vary independently and are not correlated. It also suggests that the particles with the brightest $N_2V^-$

PL show very little NV$^-$ PL and vice versa. On the other hand, we find that N$_2$V$^-$ and N$_2$V$^0$ PL are correlated in 5 μm HPHT particles (see SI Figure S11) in agreement with our observations in bulk samples.

Having established the simultaneous presence of two N$_2$V charge states in various HPHT diamond samples, we investigated N$_2$V PL in nitrogen-doped CVD diamond samples. Figure 3 i) shows PL spectra (470 nm excitation) of three CVD samples with 10-20 ppm of N$_s^0$ compared to a bulk HPHT sample known to contain N$_2$V$^0$ (HPHT, black trace). One of the CVD samples was as-grown (CVD, blue trace), and two were processed (irradiated and annealed; see SI for more details) to create NV centers (CVD, low/high NV, red and green traces). In the spectral region between 500 and 540 nm, all CVD PL spectra exhibit the same overall shape as the HPHT spectrum and show a N$_2$V$^0$ ZPL at 503 nm and an N$_2$V$^0$ phonon side band (PSB) at ~511 nm.[36] The PL of the CVD samples processed to contain NV centers sharply increases toward longer wavelengths due to NV PL (see SI Figure S12 for complete spectra). The as-grown CVD sample follows the spectra shape of the HPHT sample and only shows a weak signal at 575 nm from the NV$^0$ ZPL. These spectra demonstrate that CVD samples with N$_s^0$ concentrations on the order of 10 ppm contain N$_2$V defects before and after they were processed to create NV centers. Interestingly, we did not observe N$_2$V$^-$ PL in any of these samples using 785 nm excitation.

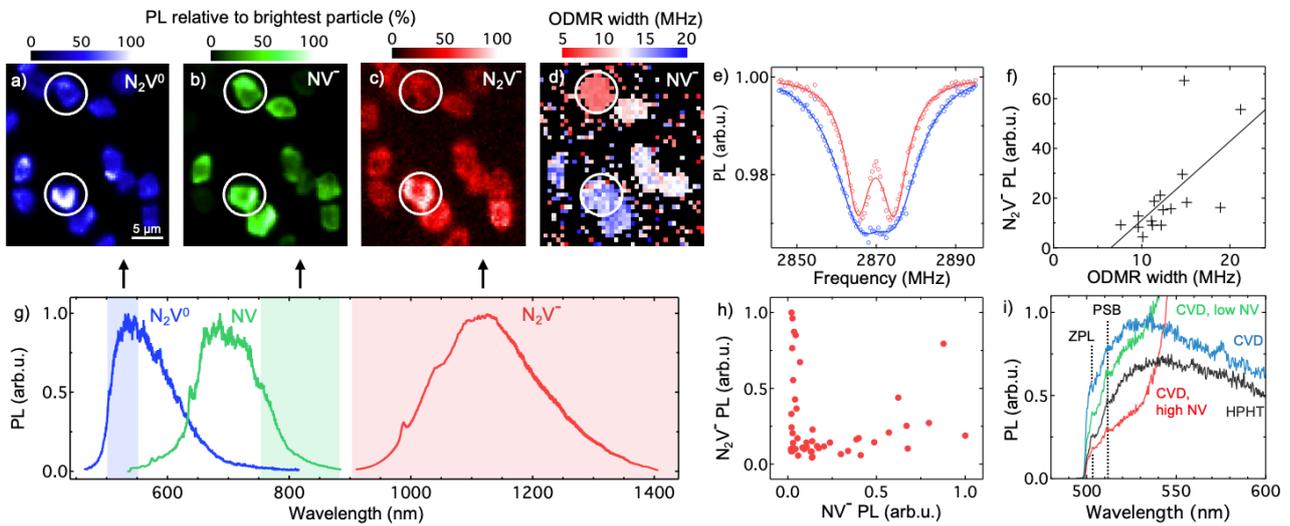

*Figure 3. Co-existence of N$_2$V and NV color centers in unprocessed HPHT and CVD diamond. a)-c) PL images of 5 μm HPHT diamond particles acquired using 440 nm (a), 520 nm (b), and 785 nm (c) excitation and by collecting PL in the spectral bands indicated in g). d) Optically detected magnetic resonance (ODMR) linewidth map of the particles shown in a)-c). The map shows the full-width half-maximum (FWHM) of the NV$^-$ ODMR peak at 2.87 GHz. e) ODMR spectra of the two particles highlighted in d) with double Lorentzian fits (solid traces) to the experimental data (circles). f) N$_2$V$^-$ PL as a function of the ODMR FWHM. Each data point represents the N$_2$V$^-$ PL intensity and ODMR width averaged over individual particles in images like those shown in panels c) and d). g) Typical PL spectra of N$_2$V$^0$, NV, and N$_2$V$^-$. Shaded areas indicate the spectral bands used to acquire images a)-c). h) Scatter plot of the N$_2$V$^-$ PL and NV$^-$ PL based on the PL from 53 individual particles. i) PL spectra of an as-grown CVD sample (blue trace) and two electron irradiated and annealed CVD samples with high and low NV concentrations (red and green traces, respectively) compared to a bulk HPHT diamond sample (black trace).*

To better understand the optical transitions underlying N$_2$V$^-$ PL, we determined its PL lifetime and analyzed cryogenic N$_2$V$^-$ PL spectra based on theoretical models of the N$_2$V$^-$ center. Figure 4 a) shows the time-resolved PL decay of N$_2$V$^-$ (red circles) and N$_2$V$^0$ (blue circles) in an unprocessed HPHT diamond sample acquired using pulsed 750 nm and 470 nm excitation, respectively. The solid lines represent single exponential fits to the decays, and black circles show the system's instrument response function (IRF). The inset shows the first nanosecond of the $N_2V^-$ decay compared to the IRF. Using deconvolution of the $N_2V$ decays and the IRF, we determine the PL lifetime of $N_2V^-$ to be $0.323 \pm 0.002$ ns. We note that this is the first time the lifetime of $N_2V^-$ is reported. The short PL lifetime agrees with PL saturation measurements, which show a very high saturation excitation intensity of $N_2V^-$ PL of $171 \pm 14$ mW (see SI Figure S13). Given that the observed lifetime ($\tau$) is given by $\tau = 1/(k_r+k_{nr})$, where $k_r$ and $k_{nr}$ are the radiative and non-radiative decay rates, respectively, a high $k_r$, a high $k_{nr}$ or a combination of both could cause the fast overall decay. Future studies will focus on determining the radiative rate and the PL quantum yield of $N_2V^-$.

We observe a PL lifetime of $N_2V^0$ of $10.79 \pm 0.04$ ns. This is shorter than the lifetime of $16.7 \pm 0.5$ ns of $N_2V^0$ determined previously in cathodoluminescence measurements using a sample with a low $N_s^0$ concentration and 1500 ppm of $N_2$ (A-centers) that was irradiated and annealed.[11] The authors note that the presence of $N_s^0$ can significantly reduce the observed lifetime. The high concentration of $N_s^0$ of 200 ppm in the sample used in our experiments likely causes the shorter lifetime observed here. Our team has previously reported that high concentrations of $N_s^0$ also shorten the lifetime of the NV center.[37]

We developed a phenomenological quantum optical model of the $N_2V^-$ center following a procedure previously reported by some members of our team for the NV center.[38] We model the $N_2V^-$ center as a multi-level open quantum system with several optical-vibronic ground states and an incoherently pumped optical excited state representing the excited band edge (see SI Figure S15 and text below the figure). We assume that the system undergoes fast vibronic decay (between adjacent ground vibronic states) and dephasing (due to the fluctuations of excited state energy relative to ground states), in addition to optical excitation and decay. The width of the zero-phonon line is assumed to be determined by the level of dephasing. In contrast, the widths of all other sidebands of the emission spectrum are affected by both dephasing and vibronic decay. We also calculated $N_2V^-$ PL spectra using complementary density functional theory (DFT). A 3D diamond nanocrystal consisting of 118 carbon atoms, an $N_2V^-$ defect at its center, and 106 hydrogen atoms terminating the crystal at the edges was studied (see SI Figure S17 for a 3D rendered geometry of the crystal). Its photophysical properties were calculated with the Orca software package[39] using the general approach reported by Karim et al.[40] PL spectra were computed using a vertical gradient approximation. See SI for more details on DFT calculations.

Figure 4 b) shows an experimental $N_2V^-$ spectrum at 5 K (red trace), a fit of the quantum optical model to the experimental data (green trace), and a PL spectrum calculated using DFT (black trace). The experimental spectrum shows a ZPL at $984.7 \pm 0.5$ nm, blue-shifted by about 1.6 nm relative to the spectral position at room temperature (300 K) of $986.5 \pm 0.5$ nm. This shift is small but agrees qualitatively with a 2.1 nm shift from 986.3 nm at 5 K to 988.4 nm at 278 K observed in another study.[19] The ZPL FWHM decreases from $7.5 \pm 0.1$ nm at 300 K to $5.4 \pm 0.1$ at 5 K. See SI Figure S14 for a direct comparison of spectra at 5 K and 300 K. Two phonon side bands (PSBs) are present at $1034.7 \pm 0.5$ nm and $1191.3 \pm 0.5$ nm, separated by 61.5 meV and $2 \times 61.5$ meV from the ZPL, in agreement with literature.[19] Comparing the phenomenological quantum optical model with the measured spectrum at 5 K, we obtain reasonable estimates for the possible PSB peaks, rates of optical decay through each emission band (the summation of which results in the total rate of 1/0.3 ns), rates of fast vibronic decay between adjacent ground levels and dephasing, as tabulated in the SI Table S2. Adjusting the dephasing rate to match the ZPL width and accounting for the temperature dependence of vibronic decay rates (assuming Bose statistics) closely reproduces the measured 300 K spectrum with no further parameter changes. Thus, we find that the phenomenological quantum optical model, together with the predicted parameters, captures the dominant physics of the $N_2V^-$ center from the ZPL up to ~1350 nm.

The DFT model also correctly predicts the existence of a ZPL in the NIR-II spectral region, accompanied by a PSB. It predicts a spectral ZPL position of about 175 nm red-shifted from the experimentally observed ZPL position and a strongly suppressed (relative to the ZPL) phonon sideband emission. This suppression is likely due to the small size of the nanocrystal (120 atoms) and its termination with hydrogen atoms. The spatial position of hydrogen atoms is fixed, which dampens lattice phonons. In Figure 4 b), the DFT PL spectrum was blue-shifted by 175 nm to coincide with the experimentally observed ZPL position. Figure 4 b) also includes a magnified PSB region (grey trace) to enable a qualitative comparison with experimental results, which reveals a good qualitative agreement with the experimentally observed PSB PL.

The evolution of the ZPL emission as a function of diamond sample size is now considered in detail with particular focus on the origin of the NIR-II PL from small (< 1 μm) HPHT particles and DNDs. Figure 4 c) shows PL spectra zoomed-in to the ZPL spectral region for HPHT diamond samples with various form factors, from a bulk sample to 500 nm particles, and for DNDs. All spectra were acquired at room temperature except the DND spectrum (red trace) and one of the 0.5 μm HPHT particles (yellow trace), which were also measured at 5 K. The bulk sample and 10 μm particles show the narrowest and most pronounced ZPL. The ZPL significantly broadens as the particle size decreases from 5 to 1 μm. The ZPL also red-shifts by about 0.6 nm for the 1 μm particles compared to the bulk sample. For 500 nm HPHT particles, a ZPL is still visible at cryogenic temperatures (5 K, yellow trace)) but disappears at room temperature (orange trace). As observed for the bulk sample in Figure 4 b), the ZPL blue-shifts by about 1.7 nm compared to bulk diamond at room temperature. For DNDs, a ZPL is not observed even at 5 K (Figure 4 c). Figure 4 d) shows the FWHM of the Gaussian fits to the ZPL shown in Figure 4 c) for the bulk diamond and 10-1 μm particles. It shows an increase in ZPL with decreasing particle size. The trend is qualitatively extrapolated with an exponential fit to the data (black trace) and suggests that the FWHM of the ZPL of diamond nanoparticles would be well above 11 nm. This strong broadening and overall weaker PL signal observed for smaller particles may explain the absence of a ZPL for the 0.5 μm HPHT particles at room temperature. It also suggests that one would not expect to observe a ZPL at any temperature in DND particles, which are more than 100 times smaller than the 0.5 μm particles, in agreement with our observations in Figure 4 c).

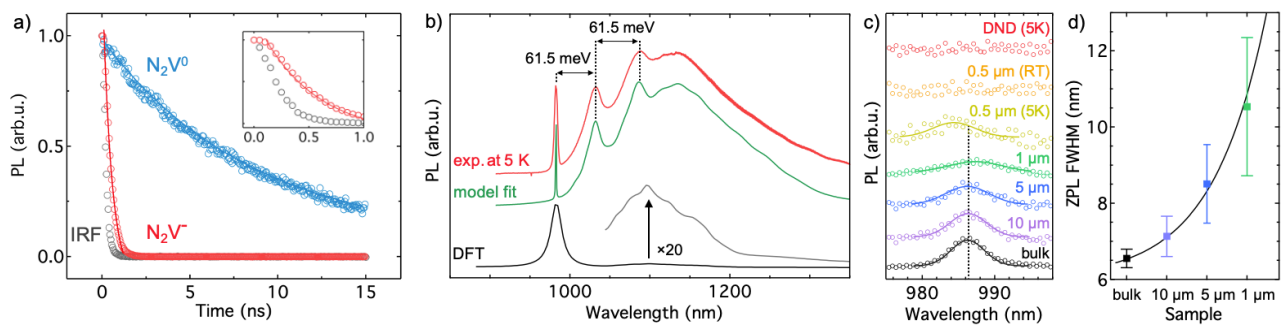

***Figure 4***. *$N_2V$ PL characterization and theoretical models of the $N_2V^-$ center. a) Time-resolved fluorescence decay measurements of $N_2V^0$ (blue circles) and $N_2V^-$ (red circles) in a bulk HPHT diamond. Solid lines represent a single exponential fit to the decay, and black circles show the instrument response function (IRF). The inset shows the first nanosecond of the decay of $N_2V^-$ compared to the IRF. b) Experimental PL spectra of the $N_2V^-$ center at 5 K (red trace) compared to spectra fitted using a quantum optical model (green trace) and a DFT model of the $N_2V^-$ PL (black trace). The grey trace shows the calculated PL spectrum magnified 20-fold to enable a comparison with experimental results. c) Evolution of the $N_2V^-$ zero phonon line (ZPL) at 968 nm for samples with different form factors from bulk diamond (black trace) to detonation nanodiamonds (red trace). Experiments for bulk, 10, 5, and 1 μm diamonds were conducted at room temperature (RT). Spectra for 0.5 μm particles acquired at room temperature (RT) and 5 K are shown, and the spectra for DNDs were acquired at 5 K. d) ZPL width as a function of diamond form factor.*

Finally, we explored NIR-II imaging of DNDs in skin cells. Briefly, DNDs were incubated with HaCaT cells (10 μg/mL) for 24 hours, followed by rinsing with phosphate-buffered saline to remove DND particles not taken up by cells. A reference sample with cells grown under the same conditions but without DND incubation was also prepared. Both cell samples were fixed and mounted on a glass coverslip for imaging. The cells were imaged using a custom-built confocal PL microscope using 785 nm excitation. See Methods for details on cell culturing and microscopy. Figure 5 a) shows a PL image of cells incubated with DNDs. To acquire this image, PL was collected in two separate spectral channels, 800 – 900 nm using a silicon-based photodiode and 900 – 1400 nm using an InGaAs photodiode, as illustrated in Figure 5 b). PL collected in the 800 – 900 nm spectral range is shown in green in the PL image in Figure 5 a), and PL collected in the 900 – 1400 nm spectral channel is shown in red. The cell nuclei, with their characteristic round to oval shape and size of 10-20 µm, and part of the cell body, can clearly be identified in the image. This PL signal likely originates from residual cell autofluorescence from molecules like lipofuscins and porphyrins, whose PL is known to extend far into the NIR-I.[41] Significantly smaller red spots (NIR-II PL) of 1-3 µm in size can be seen across the entire field of view and mostly in close vicinity of cell nuclei. Different types of nanodiamonds are known to be taken up by cells and transported to regions close to the nucleus in intracellular vesicles.[42,43] Figure 5 c) shows the autofluorescence and the signal collected in the NIR-II in the locations indicated in Figure 5 a) as a function of time under continuous 785 nm laser excitation. While the molecular autofluorescence signal decreases by 61% within the first minute of illumination, the NIR-II signal remains stable within the stability of the excitation laser beam. The NIR-II PL intensity of the small spots around the cell nucleus of ~30 k counts per second is as photostable and on the same order of magnitude as the PL observed from DND particles spin-coated onto a quartz substrate (see SI Figure S18). The reference sample of cells that were not incubated with DNDs does not show this localized PL (see SI Figure S18). Based on the above observations, we conclude that the PL from the red spots in Figure 5 a) originate from DNDs. Nanodiamonds are known to aggregate in cell media partially[44] and are also known to be present in vesicles in the cell body.[42] Hence, larger 2-3 µm sized red spots in the image in Figure 5 a) are likely larger DND aggregates. Overall, these experiments clearly demonstrate that DNDs can be used for cellular imaging in the NIR-II spectral region.

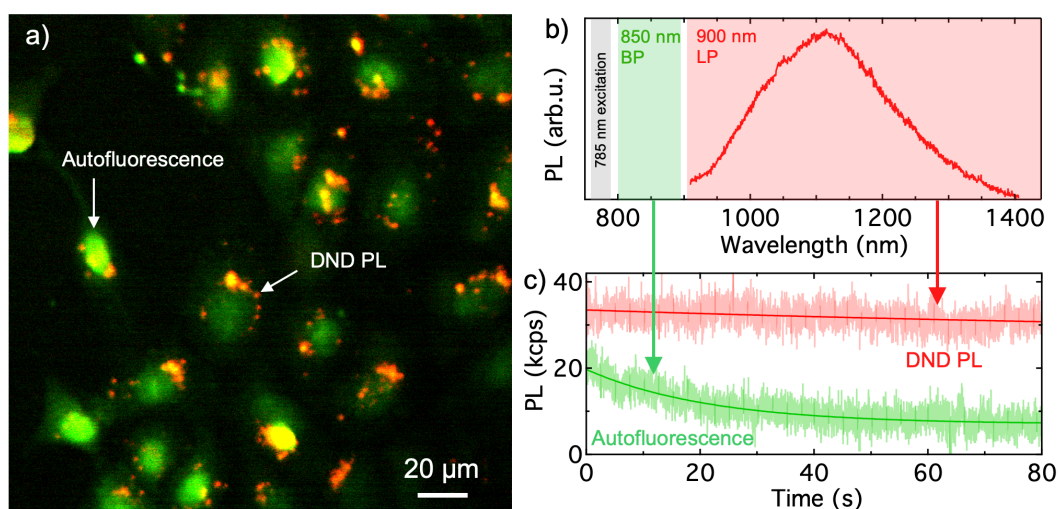

*Figure 5*. NIR-II PL imaging of detonation nanodiamonds in HaCaT skin cells. (a) Confocal PL image of fixed skin cells (HaCaT) incubated (10 µg/mL) with DNDs for 24h. b) Typical PL spectrum of DNDs. Shaded areas indicate the spectral position of the excitation laser and the optical filters used to acquire the image in a). c) Typical DND (red trace) and autofluorescence (green trace) PL intensity as a function of time acquired in the locations indicated in a). The cell autofluorescence photobleaches by 61% in the first 60 seconds of illumination; the DND PL remains stable within the drift of the excitation beam intensity (10%).

One remaining question is whether NIR-II PL from DNDs originates from the $N_2V^-$ color center or potentially from non-diamond carbon PL that is often observed for DNDs from the visible to NIR-I spectral region.[26,45] The spectral shape of the PL from DNDs is overall very similar to that of smaller HPHT diamond particles that are known to contain $N_2V^-$ centers. Based on the evolution of the ZPL width as a function of particle size shown in Figure 4 d), we do not expect to observe a ZPL in DND PL spectra. The extreme photostability shown in Figure 4 c) is commonly observed for color centers in diamond and very uncommon for non-diamond carbon or carbon-dot PL.[46] Another characteristic feature of non-diamond carbon in DNDs is that its PL spectra shift with the excitation wavelength,[45] while color center PL does not.[26] We collected DND PL spectra using 750, 700, and 650 nm excitation and find that the main spectral PL characteristics remain unchanged (see SI Figure S19). As the wavelength blue-shifts, a shoulder appears on the left side of the main DND emission peak centered at ~1100 nm, suggesting that either some non-diamond carbon PL is present or that another color center is more efficiently excited with, e.g., 650 nm light. The DND PL emission intensity decreases by about 50% upon 650 nm excitation relative to 700 nm excitation, consistent with the $N_2V^-$ absorption, which strongly decreases towards 650 nm (see Figure 25 in[9]). Lastly, DNDs are known to contain high concentrations of nitrogen > 10,000 ppm[47] and vacancies > 1,000 ppm.[48] This makes the formation of $N_2V$ centers during the detonation synthesis process feasible. It has also been reported that the concentration of the paramagnetic neutral charge state of nitrogen ($N_s^0$) is present at relatively low concentrations below 1 ppm in DNDs.[49] This suggests that most substitutional nitrogen defects donate electrons to other defects, making the creation of $N_2V^-$ via electron donation from $N_s^0$ conceivable. All of these observations strongly suggest that most of the NIR-II PL observed from DNDs originates from the $N_2V^-$ color center in diamond.

In summary, we show that the $N_2V$ defect is a ubiquitous fluorescent color center in a broad range of nitrogen-doped diamond. Both $N_2V^0$ and $N_2V^-$ are present in as-synthesized HPHT diamond from bulk samples to sub-micrometer particles in different commercial diamonds. Both the visible PL from $N_2V^0$ and the NIR-II PL from $N_2V^-$ increase with increasing substitutional nitrogen concentration in the range of 100-350 ppm in the (100) growth sector of HPHT diamonds. Both $N_2V^0$ and $N_2V^-$ can locally coexist with NV centers in micron-sized HPHT particles. The PL intensity of $N_2V^0$ and $N_2V^-$ shows a strong positive correlation in HPHT bulk samples and microparticles, but neither is correlated with NV PL intensity. $N_2V^0$ PL is observed in state-of-the-art high NV CVD diamond sensing chips. We determine a $N_2V^-$ PL lifetime of 0.3 ns and compare $N_2V^-$ PL at 5 K to a quantum optical and DFT model of the defect. We find a strong broadening of the $N_2V^-$ ZPL with decreasing particle size. Our results also suggest that $N_2V^-$ color centers are present in detonation nanodiamonds. We show that the $N_2V^-$ PL from detonation nanodiamonds can be imaged inside skin cells in the NIR-II spectral region. Overall, our results pave the way for the further exploration of the visible and NIR-II PL of the $N_2V$ center in diamond for applications from quantum memory devices to quantum sensing and imaging in biology and medicine.

**Methods**
**Materials.** Bulk diamond samples and diamond particles were obtained from commercial suppliers Element Six (UK), Sumitomo Electric Group (Japan), Pureon (Switzerland) and Taidiam Technology (China), and used as received. Detonation nanodiamonds (UD98, NanoBlox, USA) were air oxidized as described elsewhere.[50,51] See SI Table S1 for a detailed overview of samples. All particle powders were suspended in water and drop-cast or spin-coated onto quartz substrates (0.5 mm thick, Lanno Quartz, China).
**PL microscopy and spectroscopy**. Custom-built PL microscopes were used for all experiments. Standard NIR-II experiments: A continuous wave 785 nm laser beam was collimated and focused onto the sample (50× or 100×, LC Plan N IR, Olympus, Japan), PL collected through the same objective, separated from the excitation signal using a 785 nm dichroic mirror and optical 900 nm long pass filter. The PL signal was fiber-coupled and detected using an avalanche photodiode (ID230, ID Quantique, Switzerland) for imaging or a spectrometer (IsoPlane 320 fitted with a PyLoN IR detector, Teledyne, USA). Images were acquired by raster scanning the objective across the sample (PI Nano XYZ, P-545.xC7, Physik Instrumente, Germany). For time-resolved PL measurements a ps-pulsed laser was used as an excitation source (Fianium WhiteLase, NKT Photonics, Denmark) and PL analysed with a correlator card (TimeHarp 260, PicoQuant, Germany). For visible PL experiments, 440 nm – 520 nm laser excitation was used, the dichroic replaced with a 50/50 beamsplitter, PL separated from excitation laser using optical long pass filters and photons collected with a visible APD (SPCM-AQRH-TR, Excelitas, USA) or a spectrometer (IsoPlane 320 fitted with a Pixxis camera, Teledyne, USA). PL measurements at 5 K were conducted using a cryostat (Cryostation, Montana Instruments, USA).

**ODMR experiments**. NV ODMR maps were obtained using a widefield NV microscope.[34] A 532 nm laser (Gem, Novanta Photonics, USA; total input power measured at 200 mW) illuminated a region approximately 100 um in diameter. A pulsed ODMR protocol with weak microwave driving (1 μs π time) was used to avoid power broadening of the resonance linewidths. The NV fluorescence was imaged onto a sCMOS camera (Andor Zyla 5.5 USB3-W, Oxford Instruments, UK) and the spectra obtained at each pixel were fit using two Lorentzian lineshapes with shared widths.

**FTIR microscopy**. The FTIR spectra were collected using a PerkinElmer Spotlight 400 FTIR microscope. Each spectrum was measured across an area of the sample 50 × 50 μm in size. We used a reference FTIR spectrum from a CVD IIa diamond sample (Optical grade, Element Six, UK) to remove any absorption background not originating from nitrogen defects. Then, we calculated the density of $N_s^0$ defects from the absorption coefficient measured at 1130 cm$^{-1}$ as reported in the literature.[28] See SI Figure S5 for spectra and more details.

**Cell experiments**. Human skin keratinocytes (HaCaT) cells were grown in DMEM containing 10% (v/v) fetal bovine serum and 1% (v/v) penicillin-streptomycin. Cells were maintained at 37 °C, with 5% $CO_2$ and 95% relative humidity. Cells were seeded on to glass coverslips in 35mm dishes, at a seeding density of $1 \times 10^5$ cells/mL and allowed to grow for 24 h. The DND were diluted to 10 μg/mL in cell culture media and cells treated for 24 h, following which, the cells were washed in phosphate buffered saline (PBS). Cells were then fixed in 4% (v/v) formaldehyde in PBS for 10 minutes, washed thrice with PBS and mounted on to glass slides using Prolong Diamond Antifade mountant.

**Acknowledgements**
PR acknowledges support through an Australian Research Council DECRA Fellowship (DE200100279), Discover Project (DP220102518), and an RMIT University Vice-Chancellor's Senior Research Fellowship. HH acknowledges funding support through RMIT University Vice-Chancellor's Postdoctoral Research Fellowship, and the National Computational Infrastructure (NCI) supported by the Australian Government for computational resources. AM and SPR acknowledge the computational resources provided by the Australian Government through the National Computational Infrastructure National Facility and the Pawsay Supercomputer Centre. JJ and TL acknowledge funding from the German Federal Ministry for Education and Research, Bundesministerium fur Bildung und Forschung (BMBF) under grant no. 13N16485. MdV acknowledges funding through RMIT's Research Stipend Scholarship (RRSS-SC).

**References**

[1] M. Ruf, N. H. Wan, H. Choi, D. Englund, R. Hanson, *Journal of Applied Physics* **2021**, *130*, 070901.
[2] S. Pezzagna, J. Meijer, *Applied Physics Reviews* **2021**, *8*, 011308.
[3] N. Aslam, H. Zhou, E. K. Urbach, M. J. Turner, R. L. Walsworth, M. D. Lukin, H. Park, *Nat. Rev. Phys.* **2023**, *5*, 157.
[4] T. Zhang, G. Pramanik, K. Zhang, M. Gulka, L. Wang, J. Jing, F. Xu, Z. Li, Q. Wei, P. Cigler, Z. Chu, *ACS Sens.* **2021**, *6*, 2077.
[5] C. D. Clark, R. W. Ditchburn, H. B. Dyer, *Proc. R. Soc. Lond. Ser. A Math. Phys. Sci.* **1956**, *237*, 75.
[6] G. Davies, M. H. Nazaré, M. F. Hamer, *Proceedings of the Royal Society of London. Series A, Mathematical and Physical Sciences* **1976**, *351*, 245.
[7] B. L. Green, M. W. Dale, M. E. Newton, D. Fisher, *Phys. Rev. B* **2015**, *92*, 165204.

[8] S. Mukherjee, Z.-H. Zhang, D. G. Oblinsky, M. O. de Vries, B. C. Johnson, B. C. Gibson, E. L. H. Mayes, A. M. Edmonds, N. Palmer, M. L. Markham, Á. Gali, G. Thiering, A. Dalis, T. Dumm, G. D. Scholes, A. Stacey, P. Reineck, N. P. de Leon, *Nano Letters* **2023**, *23*, 2557.
[9] M. N. R. Ashfold, J. P. Goss, B. L. Green, P. W. May, M. E. Newton, C. V. Peaker, *Chem. Rev.* **2020**, *120*, 5745.
[10] J.-H. Hsu, W.-D. Su, K.-L. Yang, Y.-K. Tzeng, H.-C. Chang, *Appl. Phys. Lett.* **2011**, *98*, 193116.
[11] M. D. Crossfield, G. Davies, A. T. Collins, E. C. Lightowlers, *J. Phys. C: Solid State Phys.* **2001**, *7*, 1909.
[12] P. Udvarhelyi, G. Thiering, E. Londero, A. Gali, *Phys. Rev. B* **2017**, *96*, 155211.
[13] G. Bian, J. Zhang, L. Xu, P. Fan, M. Li, C. Wu, J. Li, H. Wang, Q. Zhang, Z. Cai, H. Yuan, *Adv. Quantum Technol.* **2022**, *5*, DOI 10.1002/qute.202200044.
[14] Y. Mita, Y. Nisida, K. Suito, A. Onodera, S. Yazu, *Journal of Physics: Condensed Matter* **1990**, *2*, 8567.
[15] P. Deák, B. Aradi, M. Kaviani, T. Frauenheim, A. Gali, *Phys. Rev. B* **2014**, *89*, 075203.
[16] N. B. Manson, M. Hedges, M. S. J. Barson, R. Ahlefeldt, M. W. Doherty, H. Abe, T. Ohshima, M. J. Sellars, *New Journal of Physics* **2018**, *20*, 113037.
[17] A. T. Collins, *Journal of Physics: Condensed Matter* **2002**, *14*, 3743.
[18] L. D. Cas, S. Zeldin, N. Nunn, M. Torelli, A. I. Shames, A. M. Zaitsev, *Advanced Functional Materials* **2019**, *29*, 1.
[19] S. C. Lawson, G. Davies, A. T. Collins, A. Mainwood, *J. Phys.: Condens. Matter* **1992**, *4*, 3439.
[20] P. R. Buerki, I. M. Reinitz, S. Muhlmeister, S. Elen, *Diam. Relat. Mater.* **1999**, *8*, 1061.
[21] C. Bradac, T. Gaebel, C. I. Pakes, J. M. Say, A. V. Zvyagin, J. R. Rabeau, *Small* **2013**, *9*, 132.
[22] S. L. Y. Chang, P. Reineck, A. Krueger, V. N. Mochalin, *Acs Nano* **2022**, *16*, 8513.
[23] K. Turcheniuk, V. N. Mochalin, *Nanotechnology* **2017**, *28*, 252001.
[24] N. Nunn, M. d'Amora, N. Prabhakar, A. M. Panich, N. Froumin, M. D. Torelli, I. Vlasov, P. Reineck, B. Gibson, J. M. Rosenholm, S. Giordani, O. Shenderova, *Methods Appl. Fluoresc.* **2018**, *6*, 035010.
[25] G. Davies, *J. Phys. C: Solid State Phys.* **1972**, *5*, 2534.
[26] P. Reineck, D. W. M. Lau, E. R. Wilson, K. Fox, M. R. Field, C. Deeleepojananan, V. N. Mochalin, B. C. Gibson, *ACS Nano* **2017**, *11*, 10924.
[27] G. Davies, *J. Phys. C: Solid State Phys.* **2001**, *9*, L537.
[28] I. Kiflawi, A. E. Mayer, P. M. Spear, J. A. V. Wyk, G. S. Woods, *Philos. Mag. Part B* **1994**, *69*, 1141.
[29] T.-L. Wee, Y.-W. Mau, C.-Y. Fang, H.-L. Hsu, C.-C. Han, H.-C. Chang, *Diam. Relat. Mater.* **2009**, *18*, 567.
[30] R. C. Burns, V. Cvetkovic, C. N. Dodge, D. J. F. Evans, M.-L. T. Rooney, P. M. Spear, C. M. Welbourn, *J. Cryst. Growth* **1990**, *104*, 257.
[31] F. A. Hahl, L. Lindner, X. Vidal, T. Luo, T. Ohshima, S. Onoda, S. Ishii, A. M. Zaitsev, M. Capelli, B. C. Gibson, A. D. Greentree, J. Jeske, *Sci. Adv.* **2022**, *8*, eabn7192.
[32] L. Lindner, F. A. Hahl, T. Luo, G. N. Antonio, X. Vidal, M. Rattunde, T. Ohshima, J. Sacher, Q. Sun, M. Capelli, B. C. Gibson, A. D. Greentree, R. Quay, J. Jeske, *Sci. Adv.* **2024**, *10*, eadj3933.
[33] J. Jeske, J. H. Cole, A. D. Greentree, *N. J. Phys.* **2016**, *18*, 013015.
[34] S. C. Scholten, A. J. Healey, I. O. Robertson, G. J. Abrahams, D. A. Broadway, J.-P. Tetienne, *Journal of Applied Physics* **2021**, *130*, 150902.
[35] E. Bauch, S. Singh, J. Lee, C. A. Hart, J. M. Schloss, M. J. Turner, J. F. Barry, L. M. Pham, N. Bar-Gill, S. F. Yelin, R. L. Walsworth, *Phys. Rev. B* **2020**, *102*, 134210.
[36] K. Iakoubovskii, G. J. Adriaenssens, N. N. Dogadkin, A. A. Shiryaev, *Diam. Relat. Mater.* **2001**, *10*, 18.
[37] M. Capelli, L. Lindner, T. Luo, J. Jeske, H. Abe, S. Onoda, T. Ohshima, B. C. Johnson, D. Simpson, A. Stacey, P. Reineck, B. Gibson, A. D. Greentree, *New Journal of Physics* **2022**.

[38] H. Hapuarachchi, F. Campaioli, J. H. Cole, *Nanophotonics* **2022**, *11*, 4919.
[39] F. Neese, *Wiley Interdiscip. Rev.: Comput. Mol. Sci.* **2012**, *2*, 73.
[40] A. Karim, I. Lyskov, S. P. Russo, A. Peruzzo, *J. Appl. Phys.* **2020**, *128*, 233102.
[41] A. C. Croce, G. Bottiroli, *Eur. J. Histochem.* **2014**, *58*, 2461.
[42] P. Reineck, A. N. Abraham, A. Poddar, R. Shukla, H. Abe, T. Ohshima, B. C. Gibson, C. Dekiwadia, J. J. Conesa, E. Pereiro, A. Gelmi, G. Bryant, *Biotechnology Journal* **2020**, *2000289*, 2000289.
[43] M. Chipaux, K. J. van der Laan, S. R. Hemelaar, M. Hasani, T. Zheng, R. Schirhagl, *Small* **2018**, *14*, 1704263.
[44] E. R. Wilson, L. M. Parker, A. Orth, N. Nunn, M. D. Torelli, O. Shenderova, B. Gibson, P. Reineck, *Nanotechnology* **2019**, *30*, 385704.
[45] P. Reineck, D. W. M. Lau, E. R. Wilson, N. Nunn, O. A. Shenderova, B. C. Gibson, *Scientific Reports* **2018**, *8*, 2.
[46] P. Reineck, A. Francis, A. Orth, D. W. M. Lau, R. D. V. Nixon-Luke, I. D. Rastogi, W. A. W. Razali, L. M. Parker, V. K. A. Sreenivasan, L. J. Brown, B. C. Gibson, *Advanced Optical Materials* **2016**, *4*, 1549.
[47] I. I. Vlasov, O. Shenderova, S. Turner, O. I. Lebedev, A. A. Basov, I. Sildos, M. Rähn, A. A. Shiryaev, G. V. Tendeloo, *Small* **2010**, *6*, 687.
[48] S. L. Chang, A. S. Barnard, C. Dwyer, C. Boothroyd, R. K. Hocking, E. Osawa, R. J. Nicholls, *Nanoscale* **2016**, *1*, 10548.
[49] O. A. Shenderova, I. I. Vlasov, S. Turner, G. V. Tendeloo, S. B. Orlinskii, A. A. Shiryaev, A. A. Khomich, S. N. Sulyanov, F. Jelezko, J. Wrachtrup, *J. Phys. Chem. C* **2011**, *115*, 14014.
[50] S. Osswald, G. Yushin, V. Mochalin, S. O. Kucheyev, Y. Gogotsi, *J. Am. Chem. Soc.* **2006**, *128*, 11635.
[51] V. N. Mochalin, O. Shenderova, D. Ho, Y. Gogotsi, *Nat. Nanotechnol.* **2012**, *7*, 11.

# Supporting Information for:

# The N$_2$V color center: a ubiquitous visible and near-infrared-II quantum emitter in nitrogen-doped diamond

Brett C. Johnson[1], Mitchell O. de Vries[1], Alexander J. Healey[1], Marco Capelli[1], Anjay Manian[1], Giannis Thalassinos[1], Amanda N. Abraham[1], Harini Hapuarachchi[1], Tingpeng Luo[2] Vadym Mochalin[3], Jan Jeske[2], Jared H. Cole[1], Salvy Russo[1], Brant C. Gibson[1], Alastair Stacey[1], Philipp Reineck[1]

[1] School of Science, RMIT University, Melbourne, VIC 3001, Australia
[2] Fraunhofer Institute for Applied Solid State Physics IAF, Tullastraße 72, 79108 Freiburg im Breisgau, Germany
[3] Department of Chemistry & Department of Materials Science & Engineering, Missouri University of Science & Technology, Rolla, Missouri 65409, USA

**Table S1**. Overview of diamond samples used in this study.

| Sample type | Supplier | Product name |
|---|---|---|
| Bulk HPHT | Element Six, UK | SC Plate Type Ib 3.0×3.0mm, 0.30mm thick, <100> |
| Bulk HPHT | Element Six, UK | DNV-B14 3.0×3.0mm, 0.5mm thick |
| Bulk HPHT | Sumitomo Electric Group, Japan | SUMICRYSTAL |
| Bulk HPHT | Taidiam Technology, China | Yellow HPHT crystal |
| ~5 µm HPHT particles | Element Six, UK | Micron+ MDA, 3-6 µm |
| ~1 µm HPHT particles | Element Six, UK | Micron+ MDA, 0.75-1.5 µm |
| ~0.5 µm HPHT particles | Element Six, UK | Micron+ MDA, 0-1 µm |
| ~10 µm HPHT particles | Pureon, Switzerland | Microdiamant MSY 8-12 µm |
| ~5 µm HPHT particles | Pureon, Switzerland | Microdiamant MSY 4-6 µm |
| ~1 µm HPHT particles | Pureon, Switzerland | Microdiamant MSY 0.75-1.25 µm |
| Bulk HPHT | Sumitomo Electric Group, Japan | SUMICRYSTAL |
| Bulk HPHT | Taidiam Technology, China | Yellow HPHT crystal |
| Detonation nanodiamonds | NanoBlox, USA | UD98 |
| Bulk CVD | Grown at the Melbourne Centre for Nanofabrication | See Figure S12 for details |

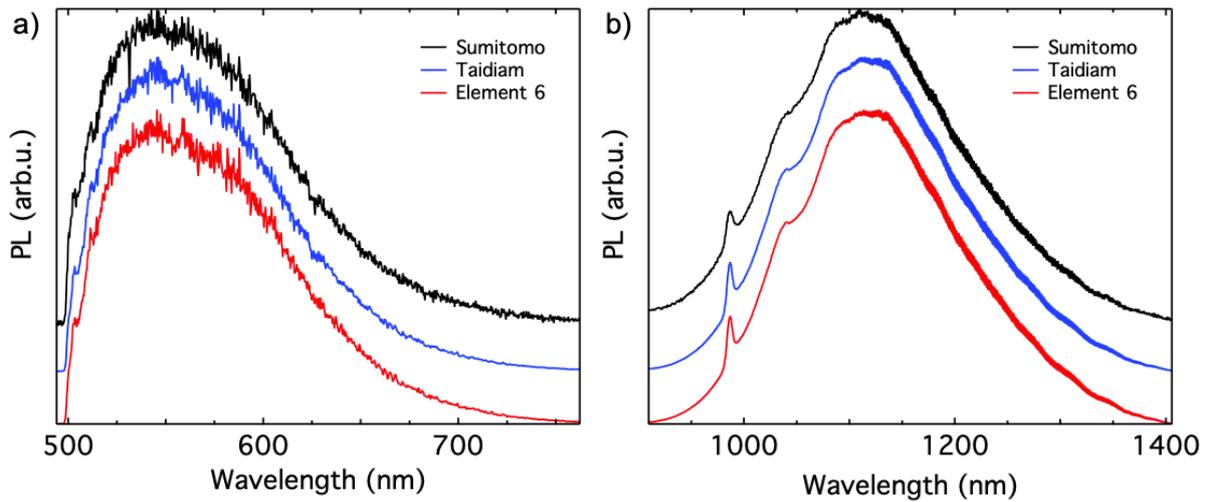

**Figure S1**. PL spectra for bulk HPHT samples from three different suppliers.

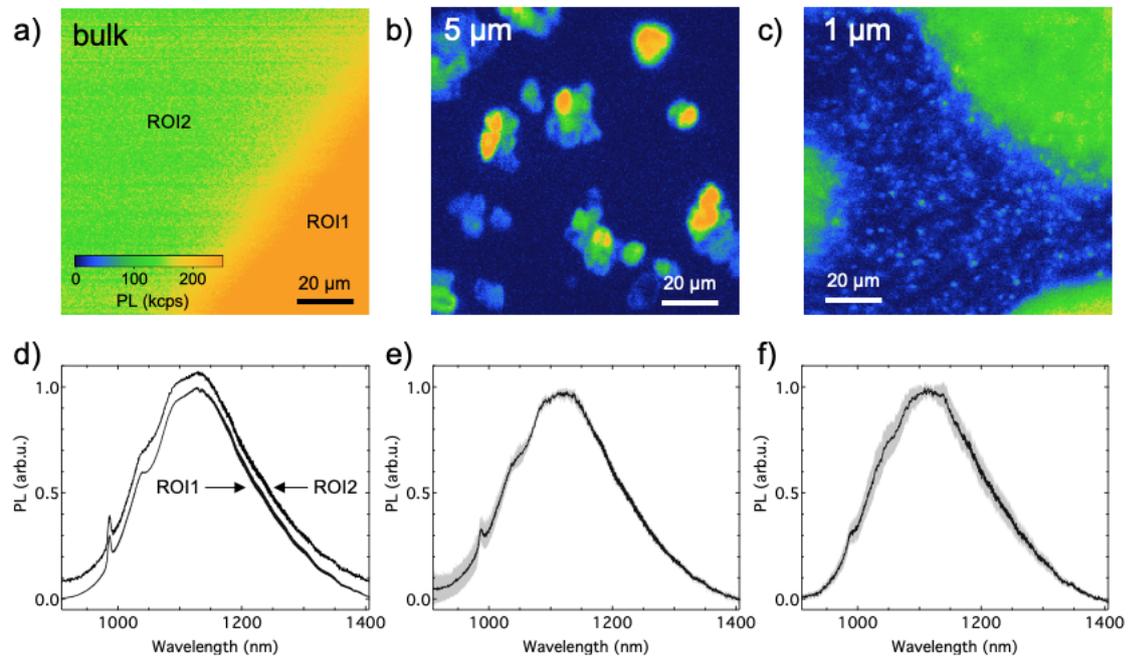

**Figure S2**. NIR-II PL images (a-c) and spectra (d-f) for bulk, 5 μm, and 1 μm HPHT diamond samples from Element Six acquired using 785 nm excitation (1 mW) and a 900 nm optical long pass filter. In the case of the particles, the PL spectra are averaged over ten different particles (solid black trace), and the grey envelope shows the standard deviation between individual particles. The PL intensity color scale in a) applies to b) and c) as well.

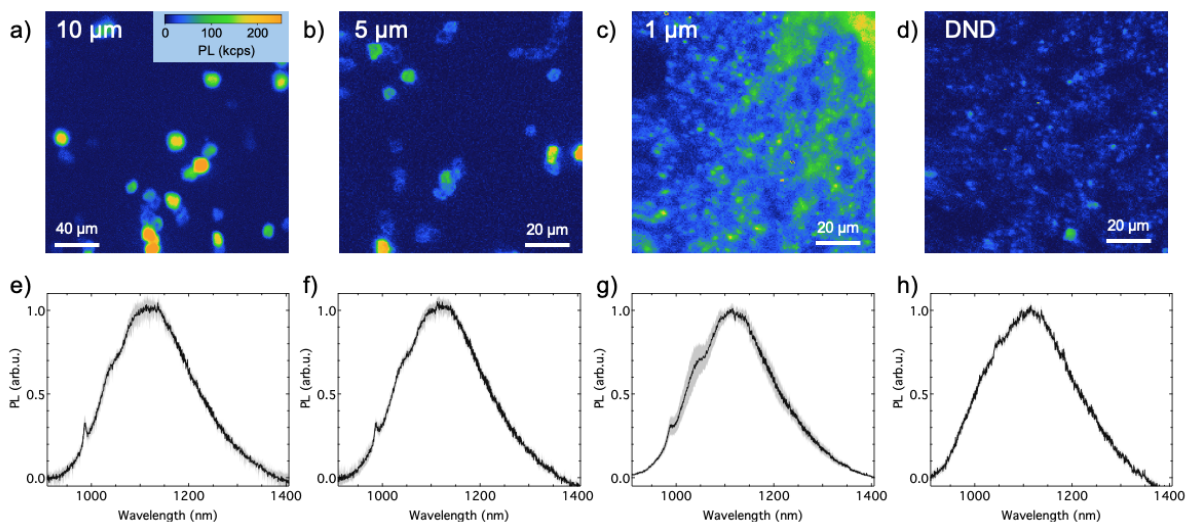

**Figure S3**. NIR-II PL images (a-d) and spectra (e-h) for all samples were acquired using 785 nm excitation (1 mW) and a 900 nm optical long pass filter. All micron-sized particles (a-c) were acquired from Pureon. The PL spectra are averaged over ten different particles (solid black trace) and the grey envelope shows the standard deviation between individual particles.

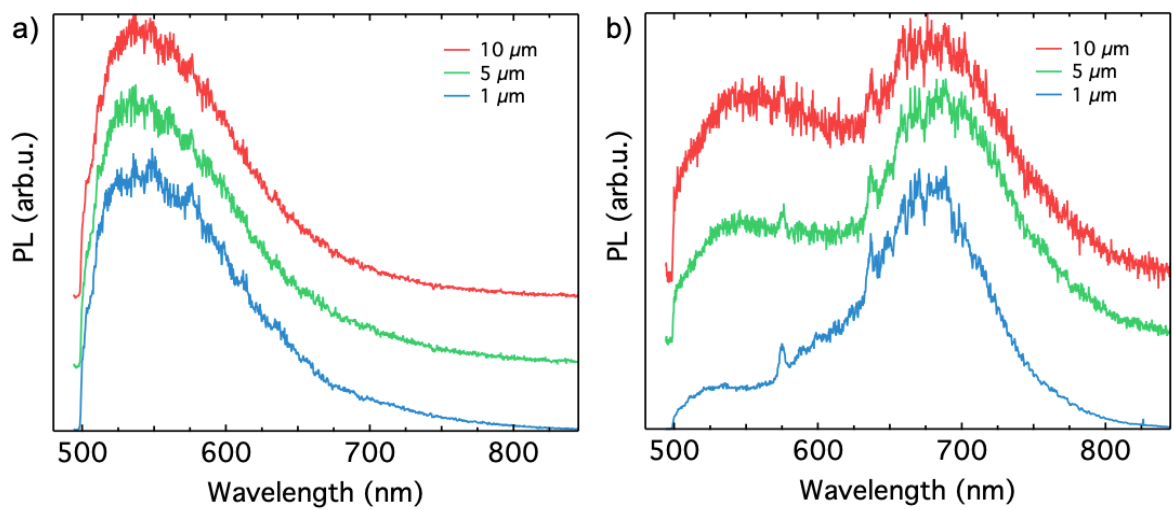

**Figure S4**. Visible PL spectra for microparticles from Pureon. Some particles show mostly $N_2V^0$ PL upon 440 nm excitation (a) but the majority show PL from both $N_2V^0$ and NV.

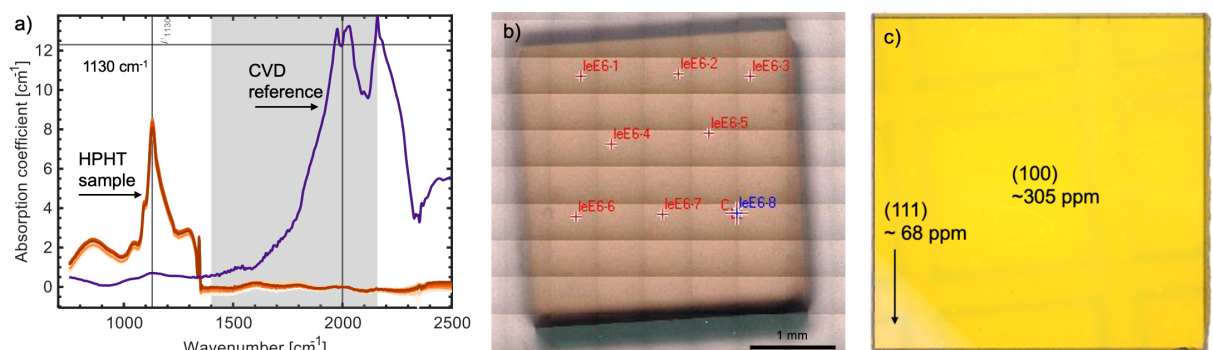

**Figure S5**. FTIR spectroscopy on bulk HPHT samples. a) FTIR spectra acquired at the locations indicated in the HPHT diamond sample in panel b). c) Bright-field image of a 3×3 mm HPHT diamond. $N_s^0$ concentrations determined via FTIR spectroscopy are indicated for the two main growth sectors.

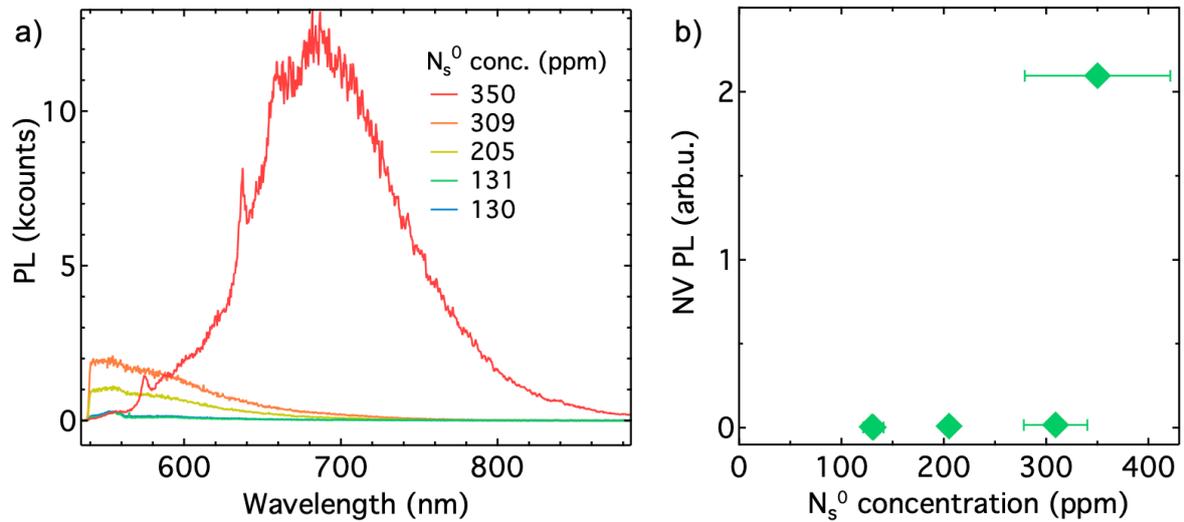

**Figure S6**. a) PL spectra of the bulk samples investigated in Figure 2 a) and 2 b) acquired using 520 nm excitation (0.6 mW). Only the sample with 350 ppm of $N_s^0$ (Taidiam) of $N_s^0$ shows NV PL. The samples with 309 and 205 ppm of $N_s^0$ (Element Six) show the tail of $N_2V^0$ PL but no NV, and the samples with 131 and 130 ppm $N_s^0$ show almost no PL. b) Integrated $NV^-$ PL from the spectra in panel a) above 700 nm.

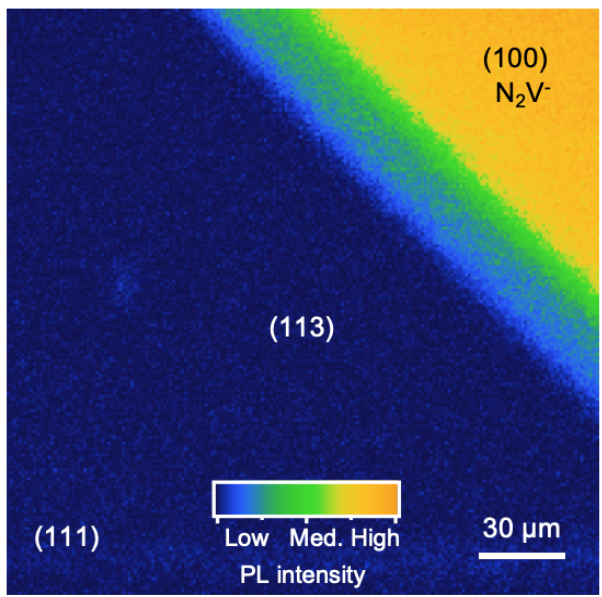

**Figure S7**. NIR-II PL image of the edge of the bulk HPHT diamond sample shown in the main text in Figure 2 d) acquired using 785 nm excitation and a 900 nm optical long pass filter.

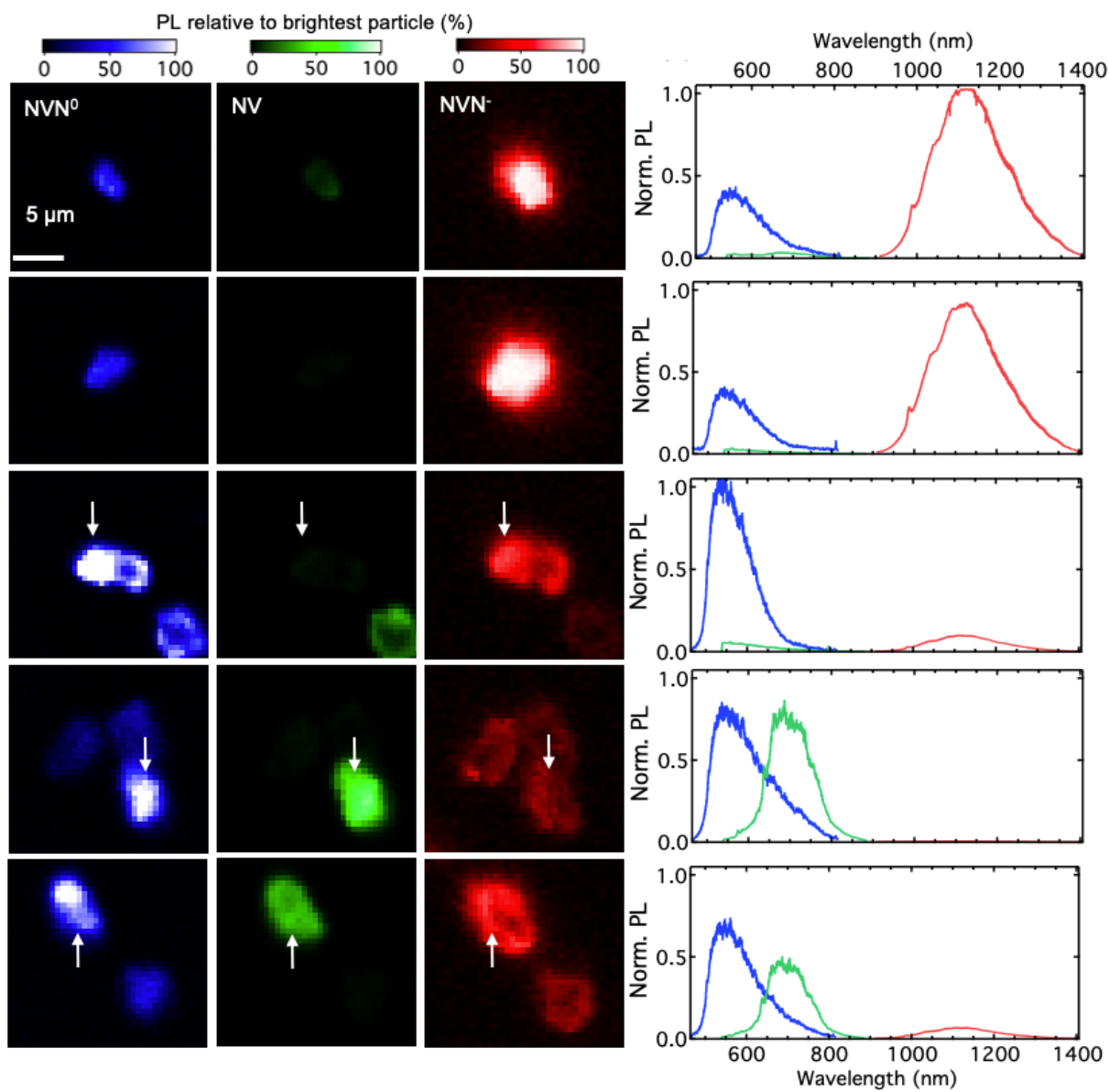

**Figure S8**. Left: PL images of 5 μm HPHT diamond particles (Pureon), acquired using 470 nm (blue images) 520 nm (green images) and 785 nm (red images) excitation and optical filters to separate the different color centers: 500-550 nm bandpass ($N_2V^0$), 700 nm long pass with a Si detector (NV), and a 900 nm long pass ($N_2V^-$). Right: Normalized PL spectra (normalized to the brightest particle in each color channel) for particles shown in images on the left. White arrows indicate the particle for which spectra are shown on the right in images that show multiple particles.

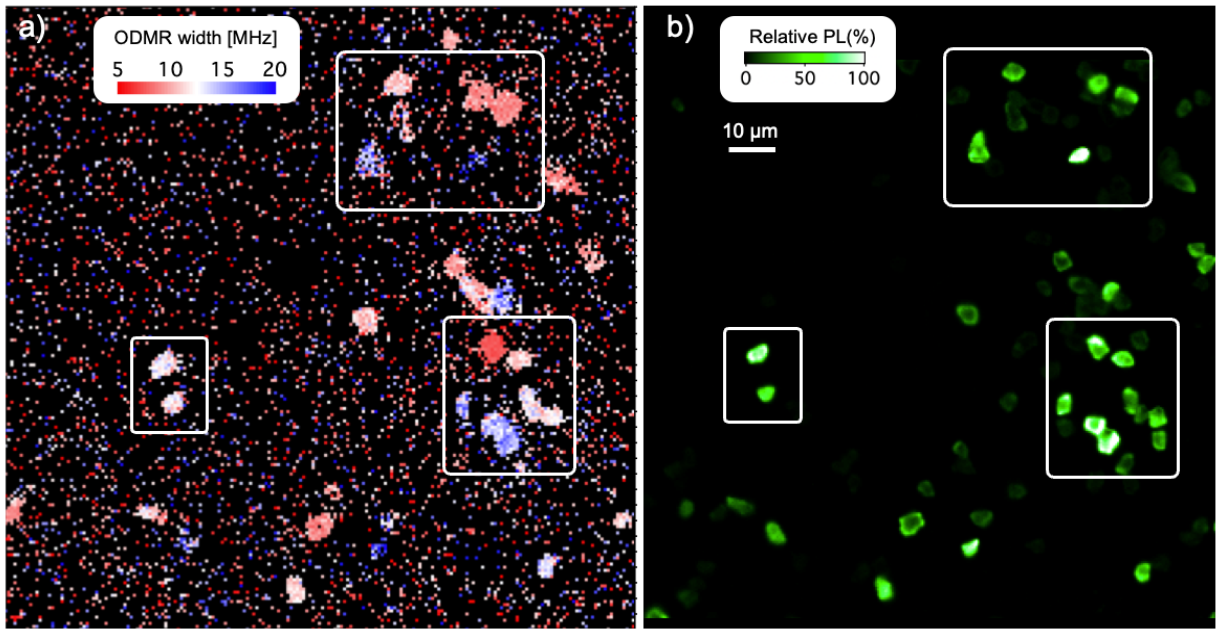

**Figure S9**. a) Map of the NV ODMR zero field splitting minimum width. b) Confocal PL image of the same region shown in a), acquired using 520 nm excitation and a 700 nm LP filter, showing the relative NV⁻ PL intensity used for ODMR measurements. Only particles with a sufficient NV⁻ PL are visible in the ODMR map.

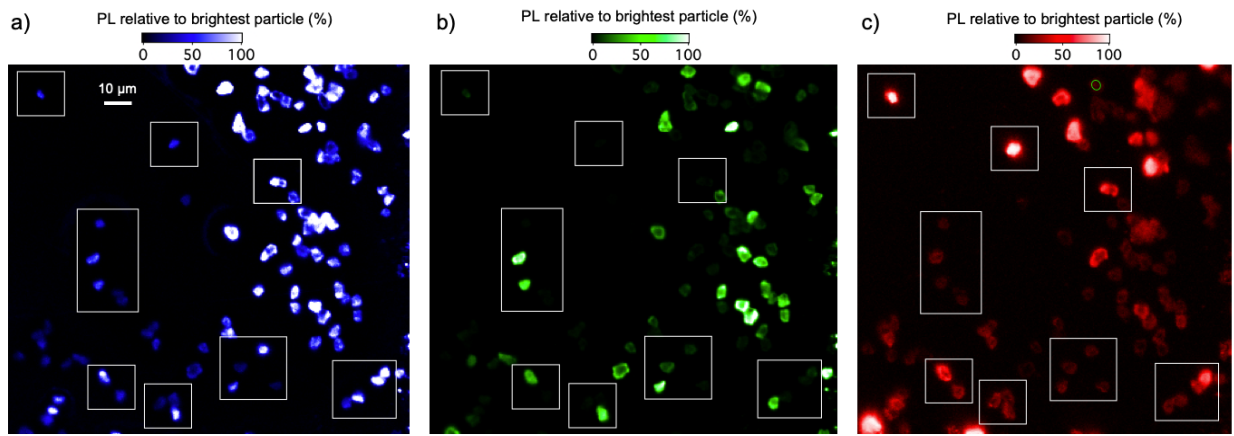

**Figure S10**. PL images of 5 μm HPHT diamond particles (Pureon), acquired using 470 nm (blue image) 520 nm (green image) and 785 nm (red images) excitation and optical filters to separate the different color centers: 500-550 nm bandpass ($N_2V^0$), 700 nm long pass (NV) with a Si detector, and a 900 nm long pass ($N_2V^-$) with an InGaAs detector.

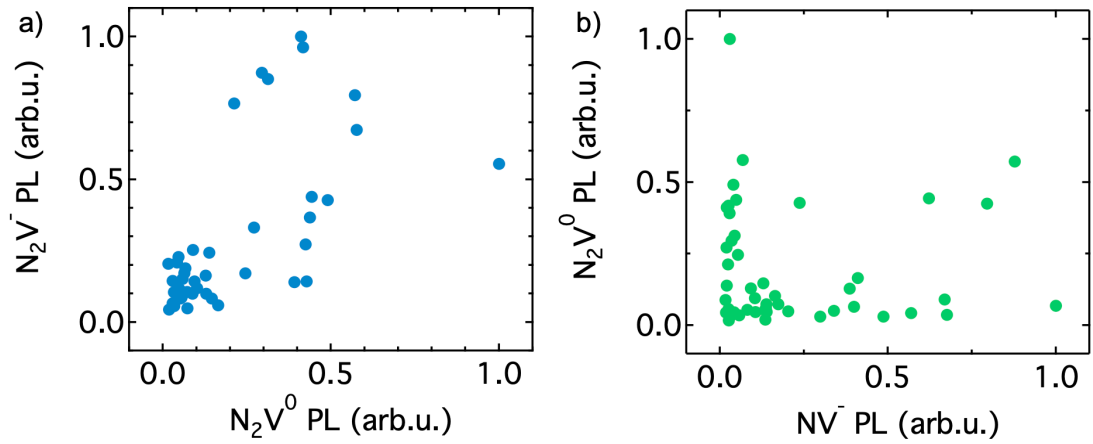

**Figure S11**. Scatter plots showing the relative PL intensities of $N_2V^-$ and $N_2V^0$ (a) and $N_2V^0$ and $NV^-$ (b) determined from 53 individual particles shown in Figure S10.

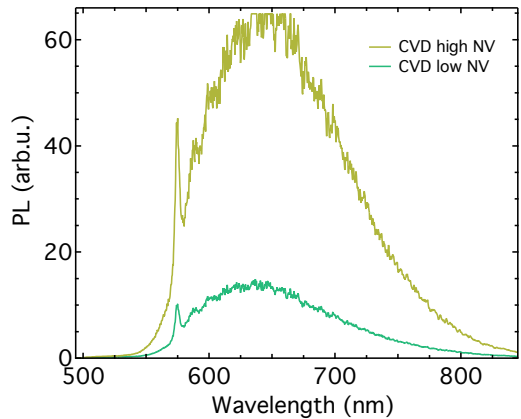

**Figure S12**. PL spectra of CVD diamond samples that were irradiated and annealed to contain high (~4.5 ppm, Element Six, DNV B14) and low (< 1 ppm) concentrations of NV centers. Spectra were acquired using 470 nm excitation (10 µW).

**CVD growth**

CVD samples made in-house were grown using plasma-enhanced chemical vapor deposition (CVD) using a CYRANNUS 2.45 GHz CVD reactor (iPlas, Germany). 300 sccm $H_2$ was flowed with 12 sccm (~4%) $CH_4$ for both growths with 5 ppm $N_2$ added for sample 'CVD' in the main text and 45 ppm $N_2$ for sample 'CVD low NV' above and in the main text. The process pressure was 160 Torr and the MW power was 3.8 kW. The substrate temperature was maintained around 900 °C for both growths using a cooling stage. In both cases, growth time was ~30 min, resulting in an N-doped layer at least 2 um thick (enough to fill the confocal volume).

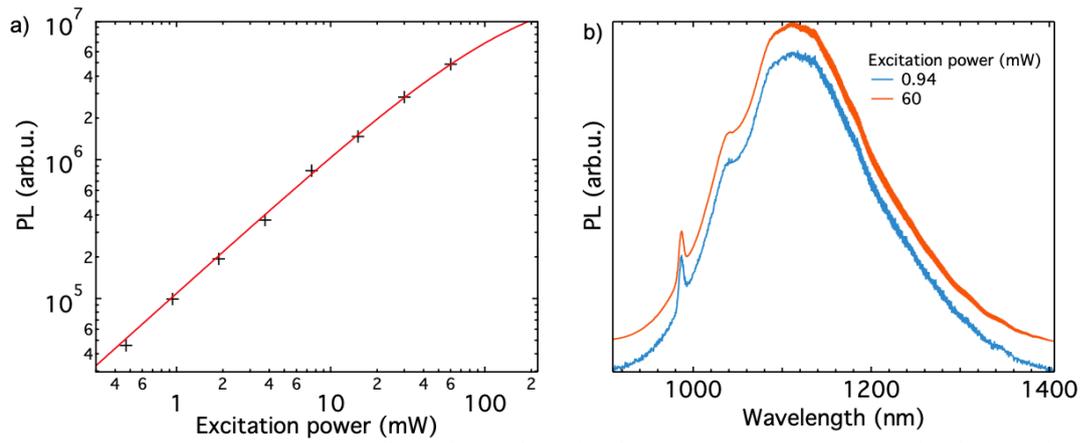

**Figure S13**. a) $N_2V^-$ PL as a function of excitation power (785 nm excitation). The red solid line is fit to the data points (black markers) using the equation $I_{PL}= I_{max} \times I_{ex} /(I_{max} + I_{sat})$, where $I_{max}$, $P_{ex}$, and $P_{sat}$ are maximum PL intensity, excitation power in mW, and saturation power in mW, respectively, yielding $P_{sat} = 171 \pm 14$ mW.

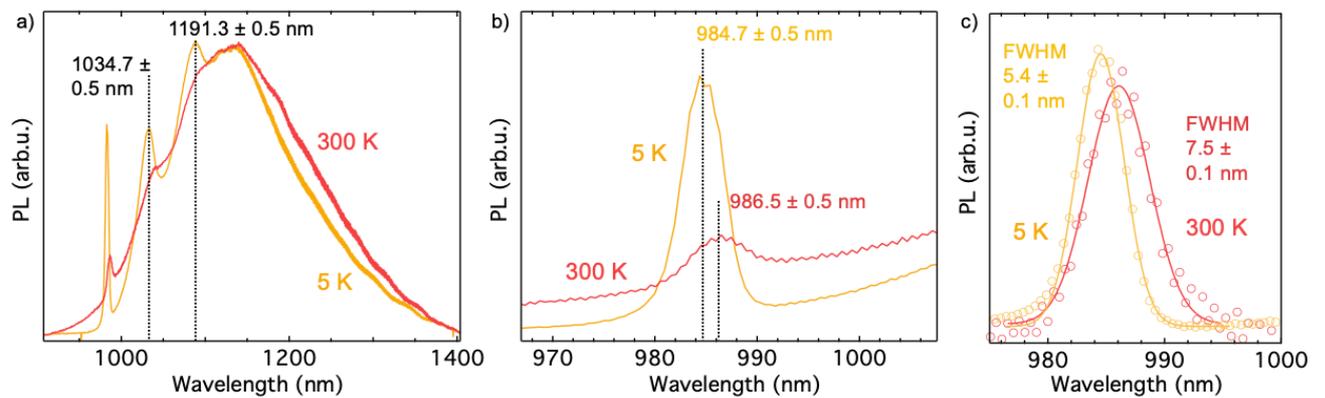

**Figure S14**. $N_2V^-$ PL spectra at 300 K (red traces) and 5K (orange traces). The spectra were acquired using 785 nm excitation and a bulk diamond HPHT sample. a) Normalized PL spectra showing the entire $N_2V^-$ spectral emission. b) Zoom into the $N_2V^-$ ZPL region. c) ZPL features at after subtraction of a polynomial background. All peak positions and spectral widths shown in the figure were determined using Gaussian fits.

# Computational Details

## Phenomenological quantum optical model for the N₂V⁻ center

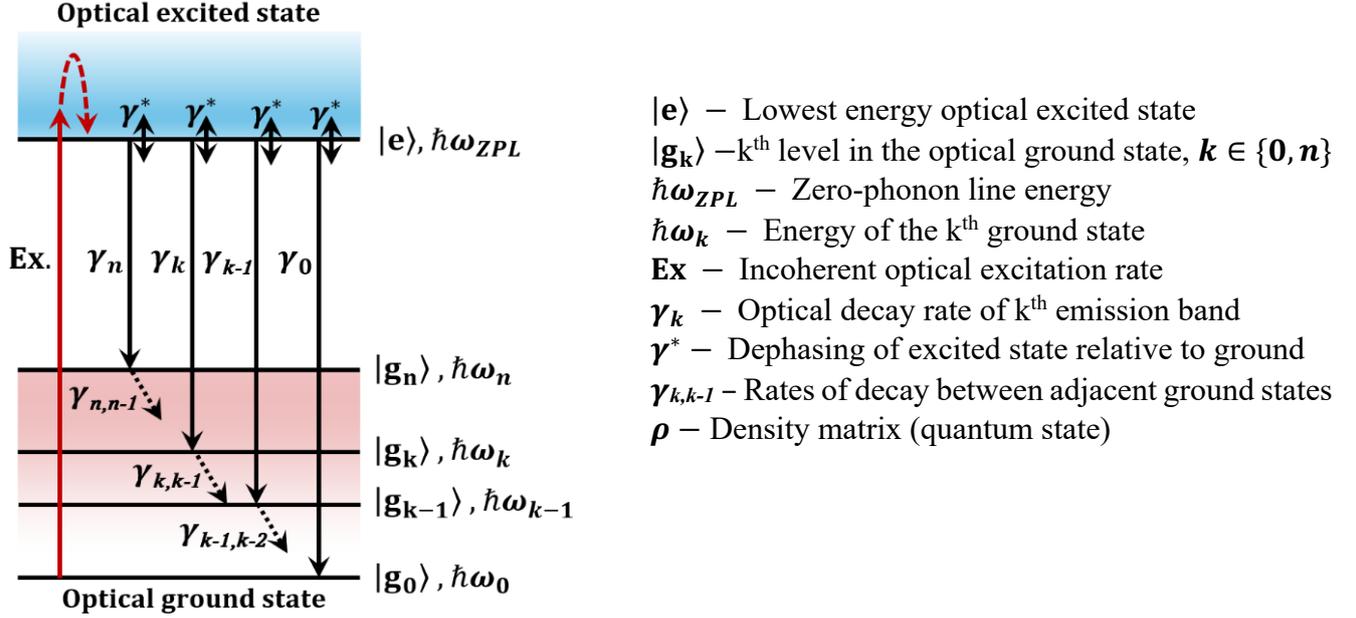

**Figure S15**. Optical states, transitions and transition rates considered in the model.

Using the modelling procedure we followed for NV⁻ centre in[1], we define the optical Hamiltonian and state evolution for the N₂V⁻ centre with the above level structure to take the following form:

**Hamiltonian:**

$$H \approx \sum_{k=0}^{n} \hbar\omega_k |g_k\rangle\langle g_k| + \hbar\omega_{ZPL}|e\rangle\langle e|$$

**Lindblad master equation-based evolution:**[2]

$$\dot{\rho}(t) \approx -\frac{i}{\hbar}[H, \rho(t)] + \sum_x \Gamma_x [L_x \rho(t) L_x^\dagger - \frac{1}{2}\{L_x^\dagger L_x, \rho(t)\}_+]$$

where $L_x$ and $\Gamma_x$ denote the incoherent processes impacting the quantum system and the corresponding rates. We include the following incoherent processes in our proposed N₂V⁻ model:

- Optical (PL) decay:
  $L_x = |g_k\rangle\langle e|$, $\Gamma_x = \gamma_k$ for $k \in \{0, ..., n\}$

- Dephasing of the optical excited state relative to ground states
  $L_x = |e\rangle\langle e|$, $\Gamma_x = \gamma^*$

- Decay between adjacent optical ground states
  $L_x = |g_{k,k-1}\rangle\langle g_k|$, $\Gamma_x = \gamma_{k,k-1}$ for $k \in \{1, ..., n\}$

We assume that the temperature dependence of these rates takes the following form, where T is the absolute temperature, $\gamma_{k,k-1}(0)$ is the spontaneous decay rate between the levels $|g_k\rangle$ and $|g_{k-1}\rangle$ at absolute zero temperature, and $k_B$ is the Boltzmann constant[3,4]:

$$\gamma_{k,k-1}(T) = \gamma_{k,k-1}(0)(1 + \bar{n}).$$

$$\bar{n} = 1/\left(\exp\left(\frac{\hbar\omega_k - \hbar\omega_{k-1}}{k_B T}\right) - 1\right).$$

- Incoherent optical excitation
  $L_x = |e\rangle\langle g_0|, \quad \Gamma_x = Ex$
  We estimate the incoherent excitation rate at 785 nm as: $Ex = P_L \sigma_{abs}/(A_L \hbar\omega_L)$

For the simulations below, we have used laser power $P_L \sim 5W$, laser spot area $A_L \sim \pi(50 \times 10^{-6})^2 m^2$. The red absorption cross-section of the $N_2V^-$ center has been estimated using that of the NV- center in [5] as $\sigma_{abs} \sim 3 \times 10^{-24} m^2$. $\hbar\omega_L$ is the photon energy at the laser wavelength 785 nm. It was verified that the simulated normalized spectral shape stays unaltered even for $P_L$ as high as 500W.

Fitting parameters:

Other model parameters are fitted to the 5 K and 300 K experimental spectra in Figure 4 of the main texts using the Lorentzian sum-based method in section V of Albrecht et al.[6]. The PL lifetime of 0.3 ns, as determined in this study, was used to fit the weighted band decay rates $\gamma_k$ below.

Table S2. Fitted model parameters.

| k | $\hbar\omega_k$ (meV) | $\gamma_k$ (MHz) | $\gamma_{k,k-1}(T=0)$ (THz) |
| --- | --- | --- | --- |
| 0 | 0.0 | 24.1 | 0.0 |
| 1 | 31.2 | 27.5 | 29.5 |
| 2 | 47.8 | 68.9 | 29.5 |
| 3 | 60.2 | 241.0 | 30.0 |
| 4 | 78.4 | 68.9 | 37.0 |
| 5 | 99.5 | 206.6 | 48.0 |
| 6 | 118.9 | 482.1 | 47.0 |
| 7 | 147.0 | 475.2 | 64.0 |
| 8 | 169.8 | 502.8 | 64.0 |
| 9 | 192.5 | 413.2 | 64.1 |
| 10 | 215.3 | 292.7 | 65.7 |
| 11 | 238.0 | 254.8 | 65.7 |
| 12 | 264.9 | 206.6 | 67.2 |
| 13 | 293.8 | 68.9 | 60.9 |

Dephasing rate at 5K ~ 1.0 (THz)
Dephasing rate at 300K ~ 15.0 (THz)

Using the above parameters and the proposed quantum optical model, $N_2V^-$ spectra at 5 K and 300 K can be theoretically reproduced as shown in Figure S16 below.

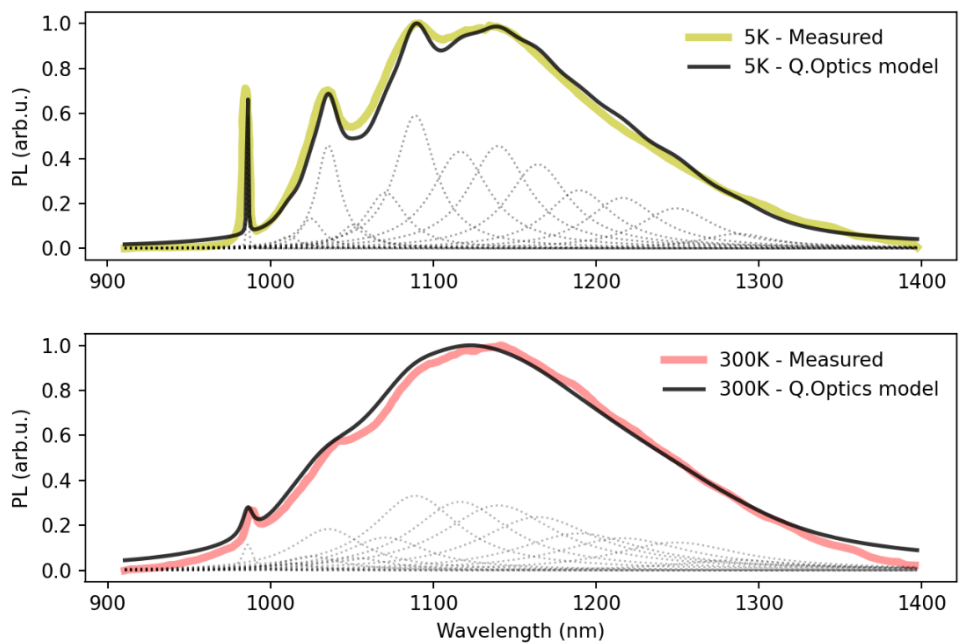

**Figure S16.** Comparison of experimental $N_2V^-$ PL spectra and spectra reproduced by the quantum optical model of $N_2V^-$.

## Density functional theory

The $N_2V$ nanocrystal was first defined so that the defect is in the center, with two carbon atoms in any direction (Figure S17). This crystal size is sufficiently large to capture essential properties representative of a bulk crystal. A smaller crystal would show much more pronounced edge effects. Significantly larger crystals would be computationally infeasible to model. The crystal is then edge-terminated with hydrogen. Ground state optimizations were then calculated and tested through frequency and electronic Hessian analysis. Following this, edge hydrogen atoms were frozen before geometries and frequencies were recalculated. This follows the approach reported by Karim and co-workers[7] except, here, hydrogen atom contributions to the electronic Hessian are also minimized. Hydrogen contamination here refers to the phenomena that C-H normal modes are typically high energy and active for nanocrystals like this; spectral calculations where they are active can result in erroneous spectral activity.

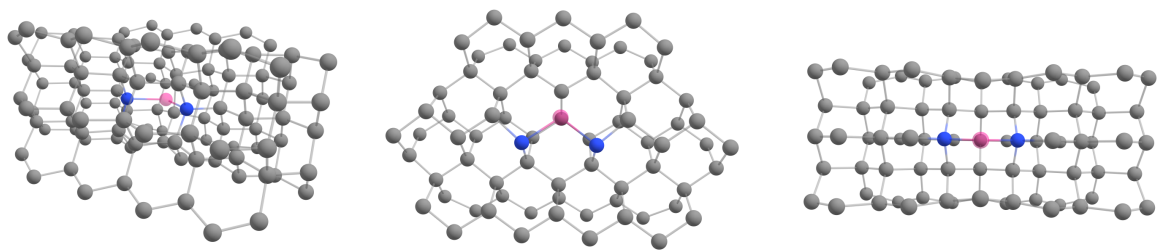

**Figure S17**. Rendered geometry of diamond nanocrystal containing a single $N_2V$ defects in its center studied in this work. The defect is centered in the nanocrystal so that there are at least two layers of carbon atoms in any direction. Grey spheres are carbon atoms, blue spheres are nitrogen, and magenta spheres are vacancies. Hydrogens were not rendered for simplicity.

Calculations are performed using the Orca software package[8], using the hybrid rendition of the Perdew-Burke-Ernzerhof (PBE0) density functional[9–11] alongside the damped variant of the Becke-Johnson geometry-dependent 3-parameter DFT-D3(BJ) dispersion correction [12,13], and the Karlsruhe variant of the single-ζ valence polarized except for hydrogen def2-SV(P) basis set[10.1039/B508541A].[14] Fluorescence spectra were generated using the vertical gradient approximation in Orca, which assumes the excited state electronic Hessian to be equivalent to the ground state electronic Hessian, and extrapolates the excited state geometry from the ground state using a displaced oscillator and augmentation method. 10 roots are examined, with fluorescence focusing on the first as per the Franck-Condon approximation.[15–17]

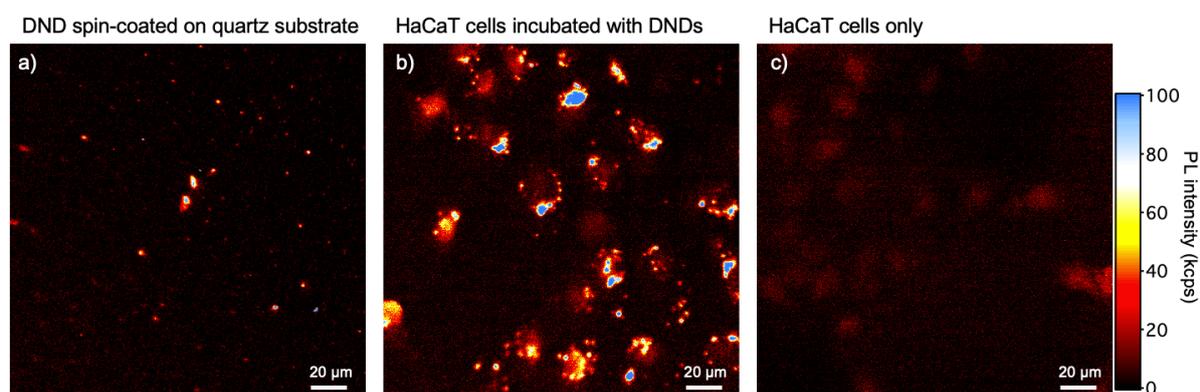

**Figure S18**. Confocal PL images of DNDs spin-coated onto a quartz substrate from an aqueous suspension (0.1 mg/mL) (a), HaCaT cells that were incubated with DNDs (10 µg/mL) for 24h (b), and a HaCat cell reference sample that was not incubated with DNDs. All images were acquired using 785 nm excitation (2 mW)

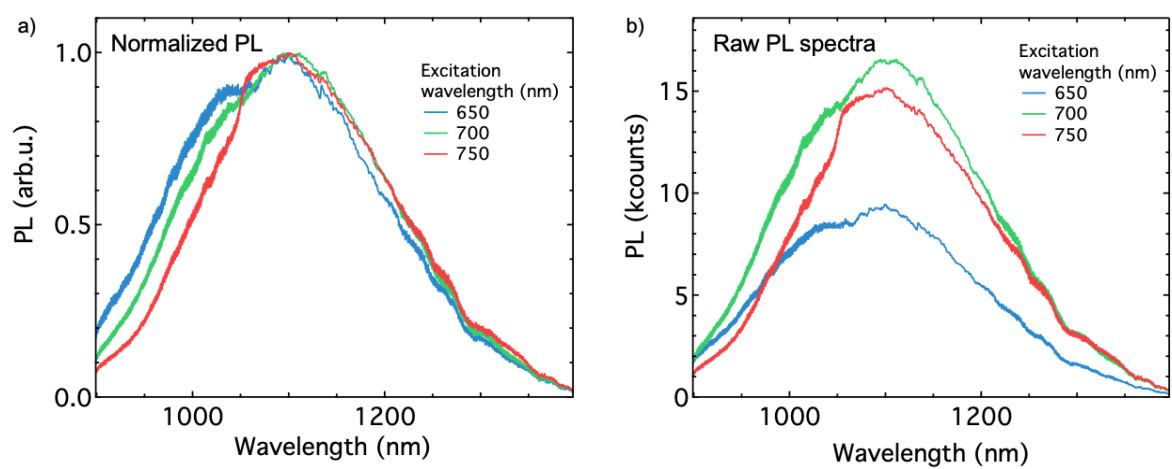

**Figure S19**. Normalized (a) and raw (PL) PL spectra of DNDs acquired using different excitation wavelengths and the same excitation intensity of 2 mW.

**References**

[1] H. Hapuarachchi, F. Campaioli, J. H. Cole, *Nanophotonics* **2022**, *11*, 4919.
[2] F. Campaioli, J. H. Cole, H. Hapuarachchi, *PRX Quantum* **2024**, *5*, 020202.
[3] W. D. Partlow, H. W. Moos, *Phys. Rev.* **1967**, *157*, 252.
[4] V. M. Acosta, *Optical Magnetometry with Nitrogen-Vacancy Centers in Diamond*, University Of California, Berkeley, **2011**.

[5] J. Jeske, D. W. M. Lau, L. P. McGuinness, P. Reineck, B. C. Johnson, J. C. McCallum, F. Jelezko, T. Volz, J. H. Cole, B. C. Gibson, A. D. Greentree, *Nature Communications* **2016**, *8*, 1.
[6] R. Albrecht, A. Bommer, C. Deutsch, J. Reichel, C. Becher, *Phys. Rev. Lett.* **2013**, *110*, 243602.
[7] A. Karim, I. Lyskov, S. P. Russo, A. Peruzzo, *J. Appl. Phys.* **2020**, *128*, 233102.
[8] F. Neese, *Wiley Interdiscip. Rev.: Comput. Mol. Sci.* **2012**, *2*, 73.
[9] J. P. Perdew, K. Burke, M. Ernzerhof, *Phys. Rev. Lett.* **1996**, *77*, 3865.
[10] C. Adamo, V. Barone, *The Journal of Chemical Physics* **1999**, *110*, 6158.
[11] M. Ernzerhof, G. E. Scuseria, *The Journal of Chemical Physics* **1999**, *110*, 5029.
[12] S. Grimme, S. Ehrlich, L. Goerigk, *Journal of Computational Chemistry* **2011**, *32*, 1456.
[13] S. Grimme, J. Antony, S. Ehrlich, H. Krieg, *The Journal of Chemical Physics* **2010**, *132*, 154104.
[14] F. Weigend, R. Ahlrichs, *Phys. Chem. Chem. Phys.* **2005**, *7*, 3297.
[15] J. Franck, E. G. Dymond, *Trans. Faraday Soc.* **1926**, *21*, 536.
[16] E. Condon, *Phys. Rev.* **1926**, *28*, 1182.
[17] E. U. Condon, *Phys. Rev.* **1928**, *32*, 858.